\documentclass[AMA,STIX1COL]{WileyNJD-v2}

\usepackage{moreverb}
\usepackage{pst-func}
\usepackage{cases}
\usepackage{tabularray}

\newcommand\BibTeX{{\rmfamily B\kern-.05em \textsc{i\kern-.025em b}\kern-.08em
T\kern-.1667em\lower.7ex\hbox{E}\kern-.125emX}
}

\articletype{Research Article}%

\received{<day> <Month>, <year>}
\revised{<day> <Month>, <year>}
\accepted{<day> <Month>, <year>}

\usepackage{boldline}

\begin{document}

\title{Local origins of quantum correlations rooted in geometric algebra}

\author[]{Joy Christian}

\authormark{Joy Christian}

\address[]{Einstein Centre for Local-Realistic Physics, Oxford OX2 6LB, United Kingdom}

\corres{Joy Christian, Einstein Centre for Local-Realistic Physics, Oxford OX2 6LB, United Kingdom\break \email{jjc@bu.edu}}

\abstract[Abstract]{In previous publications I have proposed a geometrical framework underpinning the local, realistic, and deterministic origins of the strong quantum correlations observed in Nature, without resorting to superdeterminism, retrocausality, or other conspiracy loopholes usually employed to circumvent Bell's argument against such a possibility. The geometrical framework I have proposed is based on a Clifford-algebraic interplay between the quaternionic 3-sphere, or $S^3$, which I have taken to model the geometry of the three-dimensional physical space in which we are confined to perform all our physical experiments, and an octonion-like 7-sphere, or $S^7$, which arises as an algebraic representation space of this quaternionic 3-sphere. In this paper I first review the above geometrical framework, then strengthen its Clifford-algebraic foundations employing the language of geometric algebra, and finally refute some of its critiques.}

\keywords{Clifford algebra, quantum correlations, geometric algebra, norm division algebras, local causality\break {\bf M S C Classification:} 11R52; 17C60}

\maketitle

\section{Introduction}

Unlike Einstein's theory of gravity, our other most fundamental theory of physics, namely quantum mechanics, is not a locally causal theory. This was recognized in 1935 by Einstein, Podolsky, and Rosen \cite{EPR}. They hoped, however, that quantum mechanics can be completed into a locally causal theory compatible with realism, by adding "hidden variables" to its equations. Today such a hope of restoring local causality is widely believed to have been undermined by an argument proposed by Bell \cite{Bell-1964}, which claims that no locally causal theory can reproduce all of the statistical predictions of quantum mechanics, such as those of the stronger-than-classical correlations observed in Nature \cite{Aspect}. However, I have argued elsewhere that Bell's argument is of limited validity \cite{IJTP,RSOS,IEEE-1,IEEE-2}. It depends on several implicit and explicit assumptions, which can be relaxed to achieve what Einstein had hoped for \cite{Pais}. Following Einstein's vision, I have proposed a geometrical framework underpinning the local, realistic, and deterministic origins of quantum correlations \cite{IJTP,RSOS,IEEE-1,IEEE-2}, without resorting to superdeterminism, retrocausality, or any other conspiracy loopholes usually employed to overcome Bell's argument. The geometrical framework I have proposed is based on a Clifford-algebraic interplay between the quaternionic 3-sphere, or $S^3$, which is assumed to model the geometry of the three-dimensional physical space in which we are confined to perform our physical experiments, and an octonion-like 7-sphere, or $S^7$, which arises as an algebraic representation space of this quaternionic 3-sphere. In what follows, I first review the above geometrical framework, then strengthen its Clifford-algebraic foundations using the language of geometric algebra \cite{Clifford}, and finally refute some of its critiques.

\section{Review of the quaternionic 3-sphere model of strong correlations} \label{Sec2}

It is well known that the best locally causal theory in physics is Einstein's theory of gravity, also known as the general theory of relativity. Indeed, one of the motivations of Einstein behind constructing the theory was to overcome the action-at-a-distance intrinsic to Newton's theory of gravity \cite{Newton-Cartan}. It is therefore natural to consider one of the well known cosmological solutions of Einstein's field equations of general relativity, namely, the Friedmann-Lema\^itre-Robertson-Walker solution, as our starting point:
\begin{equation}
ds^2=dt^2-a^2(t)\,d{\boldsymbol\Sigma}^2, \;\;\,d\Sigma^2=\left[\frac{d\rho^2}{1-\kappa\,\rho^2}+\rho^2 d{\boldsymbol\Omega}^2\right]\!. \label{frw}
\end{equation}
Here ${a(t)}$ is the scale factor, ${\boldsymbol\Sigma}$ is a spacelike hypersurface, ${\rho}$ is the radial coordinate within ${\boldsymbol\Sigma}$, ${\kappa}$ is the ``normalized'' curvature of ${\boldsymbol\Sigma}$, and ${\boldsymbol\Omega}$ is a solid angle within ${\boldsymbol\Sigma}$ \cite{d'Inverno}. Since we are primarily concerned with terrestrial experiments \cite{Aspect} designed to verify quantum correlations, in what follows, without loss of generality, I will restrict considerations to the current epoch of the cosmos by setting the cosmological scale factor to unity in the above line element: ${a(t) = 1}$. It still allows three possible geometries for spacetime with the product topology ${\mathrm{I\!R}\times{\boldsymbol\Sigma}}$, so that the corresponding spacelike hypersurfaces ${\boldsymbol\Sigma}$ can be isomorphic to ${\mathrm{I\!R}^3}$, $S^3$, or $H^3$, with $H^3$ being a hyperboloid of negative curvature. Among these possible geometries, only $S^3$ represents a closed universe with compact geometry and constant positive curvature. Moreover, observationally the cosmic microwave background spectra mapped by the space observatory {\it Planck} now prefers a positive curvature or $S^3$ at more than 99\% confidence level \cite{closed,Handley}.

Incidentally, Bell seems to have anticipated using gravity to negate his argument. In Chapter 7 of his book \cite{Speakable}, while exploring possible escape routes to do so, he wrote: ``The space time structure has been taken as given here. How then about gravitation?''

\subsection{Special Theorem} \label{Sec-2.1}

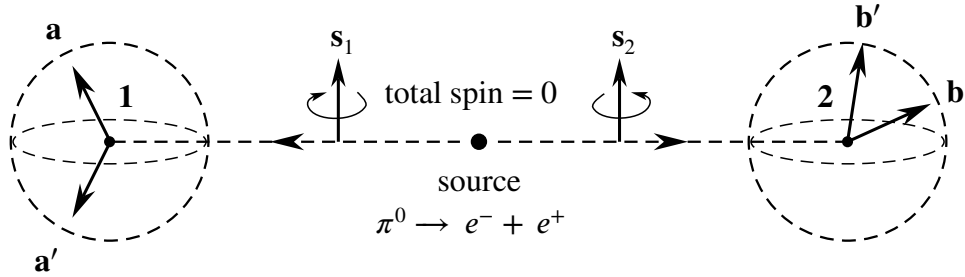
\begin{figure*}[t]
\scalebox{1}{
\begin{pspicture}(-6.6,-1.6)(4.2,1.8)

\psline[linewidth=0.1mm,dotsize=3pt 4]{*-}(-2.51,0)(-2.5,0)

\psline[linewidth=0.1mm,dotsize=3pt 4]{*-}(7.2,0)(7.15,0)

\psline[linewidth=0.4mm,arrowinset=0.3,arrowsize=3pt 3,arrowlength=2]{->}(-2.5,0)(-3,1)

\psline[linewidth=0.4mm,arrowinset=0.3,arrowsize=3pt 3,arrowlength=2]{->}(-2.5,0)(-3,-1)

\psline[linewidth=0.4mm,arrowinset=0.3,arrowsize=3pt 3,arrowlength=2]{->}(7.2,0)(8.3,0.5)

\psline[linewidth=0.4mm,arrowinset=0.3,arrowsize=3pt 3,arrowlength=2]{->}(7.2,0)(7.4,1.3)

\psline[linewidth=0.4mm,arrowinset=0.3,arrowsize=2pt 3,arrowlength=2]{->}(4.2,0)(4.2,1.1)

\psline[linewidth=0.4mm,arrowinset=0.3,arrowsize=2pt 3,arrowlength=2]{->}(0.5,0)(0.5,1.1)

\pscurve[linewidth=0.2mm,arrowinset=0.2,arrowsize=2pt 2,arrowlength=2]{->}(4.0,0.63)(3.85,0.45)(4.6,0.5)(4.35,0.65)

\put(4.1,1.25){{\large ${{\bf s}_2}$}}

\pscurve[linewidth=0.2mm,arrowinset=0.2,arrowsize=2pt 2,arrowlength=2]{<-}(0.35,0.65)(0.1,0.47)(0.86,0.47)(0.75,0.65)

\put(0.4,1.25){{\large ${{\bf s}_1}$}}

\put(-2.4,+0.45){{\large ${\bf 1}$}}

\put(6.8,+0.45){{\large ${\bf 2}$}}

\put(-3.35,1.35){{\large ${\bf a}$}}

\put(-3.5,-1.7){{\large ${\bf a'}$}}

\put(8.5,0.52){{\large ${\bf b}$}}

\put(7.3,1.5){{\large ${\bf b'}$}}

\put(1.8,-0.65){\large source}

\put(0.99,-1.2){\large ${\pi^0\longrightarrow\,e^{-}+\,e^{+}\,}$}

\put(1.11,0.5){\large total spin = 0}

\psline[linewidth=0.3mm,linestyle=dashed](-2.47,0)(2.1,0)

\psline[linewidth=0.4mm,arrowinset=0.3,arrowsize=3pt 3,arrowlength=2]{->}(-0.3,0)(-0.4,0)

\psline[linewidth=0.3mm,linestyle=dashed](2.6,0)(7.2,0)

\psline[linewidth=0.4mm,arrowinset=0.3,arrowsize=3pt 3,arrowlength=2]{->}(5.0,0)(5.1,0)

\psline[linewidth=0.1mm,dotsize=5pt 4]{*-}(2.35,0)(2.4,0)

\pscircle[linewidth=0.3mm,linestyle=dashed](7.2,0){1.3}

\psellipse[linewidth=0.2mm,linestyle=dashed](7.2,0)(1.28,0.3)

\pscircle[linewidth=0.3mm,linestyle=dashed](-2.51,0){1.3}

\psellipse[linewidth=0.2mm,linestyle=dashed](-2.51,0)(1.28,0.3)

\end{pspicture}}
\caption{A spin-less neutral pion decays into an electron-positron pair. Measurements of spin components of each separated fermion are performed at remote stations ${\mathbf{1}}$ and ${\mathbf{2}}$, obtaining binary results $\mathscr{A}=\pm1$ and $\mathscr{B}=\pm1$ along directions ${\mathbf a}$ and ${\mathbf b}$. The conservation of spin momentum dictates that the total spin of the system remains zero during its free evolution. After Christian \cite{IEEE-1}.}
\label{Fig-2}
\end{figure*}

Considering the above properties of $S^3$, in Christian \cite{IJTP,RSOS,IEEE-1,IEEE-2} I have proposed the following experimentally falsifiable \cite{IJTP} hypothesis:
\begin{hypothesis} \label{hyp1}
The strong correlations we observe in Nature \cite{Aspect} can be understood as local, realistic, and deterministic correlations if we model the three-dimensional physical space ${\boldsymbol\Sigma}$ as a closed and compact quaternionic 3-sphere using geometric algebra \cite{Clifford} rather than as a flat and open space ${\mathrm{I\!R}^3}$ using ``vector algebra.''
\end{hypothesis}
This hypothesis has far reaching consequences \cite{IJTP,RSOS,IEEE-1,IEEE-2}. To explore them, let us begin by defining the quaternionic 3-sphere as
\begin{equation}
S^3:=\left\{\,{\bf q}(\theta,\,{\mathbf r})=\varrho_r\left[\cos\left(\frac{\theta}{2}\right)+{\mathbf J}({\mathbf r})\,\sin\left(\frac{\theta}{2}\right)\right]
\Bigg|\;\left|\left|\,{\bf q}\left(\theta,\,{\mathbf r}\right)\,\right|\right|=\varrho_r\right\}\!, \label{nonsin}
\end{equation}
where ${{\mathbf J}({\mathbf r})=I_3{\mathbf r}}$ is a unit bivector (or pure quaternion) rotating about an axis vector ${{\bf r}\in{\mathrm{I\!R}}^3}$ with rotation angle ${0\leq\theta < 4\pi}$, $\varrho_r$ is the radius of the corresponding 3-sphere, and $I_3={\mathbf e}_x{\mathbf e}_y{\mathbf e}_z$ is the standard trivector. Now, for our purposes it is convenient to express the unit quaternions in (\ref{nonsin}) as products ${\bf D}({\bf n})\,{\bf L}({\bf s})$ of two bivectors, with ${\bf D}({\bf n})$ representing detectors used in Bell-test experiments \cite{Aspect} and ${\bf L}({\bf s})$ representing the spins being detected. To that end, consider the even subalgebra of the algebra $\mathrm{Cl}_{3,0}$ of orthogonal directions in ${\mathrm{I\!R}^3}$, also referred to as a bivector subalgebra \cite{Clifford}. Since the spins ${{\mathbf L}({\mathbf s}_1)}$ and ${{\mathbf L}({\mathbf s}_2)}$ and the detectors ${{\mathbf D}({\mathbf a})}$ and ${{\mathbf D}({\mathbf b})}$ are different physical systems, the experimenters, traditionally referred to as Alice and Bob, have no knowledge of the handedness of the spins until their measurements. It is therefore important to represent them using two different basis bivectors:
\begin{equation}
L_{i}\,L_{j} =-\,\delta_{ij}\,-\sum_k\epsilon_{ijk}\,L_{k} \label{bi-1-m}
\end{equation}
and
\begin{equation}
D_{i}\,D_{j}=-\,\delta_{ij}\,-\sum_k\epsilon_{ijk}\,D_{k}\,. \label{bi-2-m}
\end{equation}
These bases are then related by the orientation (or handedness) $\lambda=\pm1$ of $S^3$ as
\begin{equation}
L_{i}=\lambda\,D_{i}\;\;\Longleftrightarrow\;\;
D_{i}=\lambda\,L_{i}\,, \label{birel-m}
\end{equation}
so that, about any unit vector $\mathbf{n}$, the rotations of the spin and detector bivectors are related only by the orientation $\lambda$ of $S^3$ as
\begin{equation}
{\mathbf L}({\mathbf n})\,=\,\lambda\,{\mathbf D}({\mathbf n})\,\,\Longleftrightarrow\,\,{\mathbf D}({\mathbf n})\,=\,\lambda\,{\mathbf L}({\mathbf n})\,. \label{20no}
\end{equation}
The orientation $\lambda$, with $\lambda^2=1$, thus plays a role of a random ``hidden variable'' in the model, with 50/50 chance of $\lambda=+1$ or $\lambda=-1$ during the detection processes at the two ends of a Bell-test experiment. Since the spins --- which originate with the common cause $\lambda$ in the overlap of the backward light-cones of Alice and Bob --- are being measured with respect to the detectors at a later time, we may write them as functions of the orientation $\lambda$ to remind us of this fact: ${{\mathbf L}({\mathbf s}_1,\lambda)}$ and ${{\mathbf L}({\mathbf s}_2,\lambda)}$. In the manner of Bell \cite{Bell-1964}, the detection processes of Alice and Bob can then be specified as limiting scalar points of two quaternions within $S^3$:
\begin{align}
S^3\ni{\mathscr A}({\mathbf a},{\lambda})\,&=\lim_{{\mathbf s}_1\,\rightarrow\,{\mathbf a}}\left\{-\,{\mathbf D}({\mathbf a})\,{\mathbf L}({\mathbf s}_1,\,\lambda)\right\} =\lim_{{\mathbf s}_1\,\rightarrow\,{\mathbf a}}\left\{\,+\,{\mathbf q}(\eta_{{\mathbf a}{\mathbf s}_1},\,{\mathbf r}_1)\right\} \label{a-q} \\
&=+\lambda\lim_{{\mathbf s}_1\,\rightarrow\,{\mathbf a}}\left\{ \cos(\eta_{{\mathbf a}{\mathbf s}_1})+(I_3{\mathbf r}_{1})\sin(\eta_{{\mathbf a}{\mathbf s}_1})\right\},
\end{align}
where $\eta_{{\mathbf a}{\mathbf s}_1}$ is the angle between the spin direction ${\mathbf s}_1$ and the detector direction ${\mathbf a}$ chosen by Alice at a later time (cf. Figure~\ref{Fig-2}\!\!\!), and 
\begin{equation}
{\mathbf r}_1=\frac{{\mathbf a}\times{\mathbf s}_1}{||{\mathbf a}\times{\mathbf s}_1||} \label{rot1}
\end{equation}
is the rotation axis of the quaternion ${\mathbf q}(\eta_{{\mathbf a}{\mathbf s}_1},\,{\mathbf r}_1)$ in $S^3$. Thus, as ${\mathbf s}_1\,\rightarrow\,{\mathbf a}$ during the detection process so that the angle $\eta_{{\mathbf a}{\mathbf s}_1}\rightarrow\,0$,
\begin{equation}
{\mathscr A}({\mathbf a},{\lambda})\longrightarrow\pm1, \,\;\text{with its average}\;\Bigl\langle{\mathscr A}({\mathbf a},{\lambda})\Bigr\rangle\longrightarrow 0,
\end{equation}
because $\lambda=\pm1$ is a fair coin. Note that the rotation angle ${\theta}$ of the quaternion is twice the angle ${\eta_{{\mathbf a}{\mathbf s}_1}}$ between ${\mathbf a}$ and ${\mathbf s}_1$. Similarly, 
\begin{align}
S^3\ni{\mathscr B}&({\mathbf b},\,{\lambda})\,=\lim_{{\mathbf s}_2\,\rightarrow\,{\mathbf b}}\left\{+\,{\mathbf L}({\mathbf s}_2,\,\lambda)\,{\mathbf D}({\mathbf b})\right\} =\lim_{{\mathbf s}_2\,\rightarrow\,{\mathbf b}}\left\{\,-\,{\mathbf q}(\eta_{{\mathbf s}_2{\mathbf b}},\,{\mathbf r}_2)\right\} \label{b-q} \\
&=-\lambda\lim_{{\mathbf s}_2\,\rightarrow\,{\mathbf b}}\left\{ \cos(\eta_{{\mathbf s}_2{\mathbf b}})+(I_3{\mathbf r}_{2})\sin(\eta_{{\mathbf s}_2{\mathbf b}})\right\},
\end{align}
where $\eta_{{\mathbf s}_2{\mathbf b}}$ is the angle between the spin direction ${\mathbf s}_2$ and the detector direction ${\mathbf b}$ chosen by Bob at a later time (cf. Figure~\ref{Fig-2}\!\!\!), and 
\begin{equation}
{\mathbf r}_2=\frac{{\mathbf s}_2\times{\mathbf b}}{||{\mathbf s}_2\times{\mathbf b}||} \label{rot2}
\end{equation}
is the rotation axis of the quaternion ${\mathbf q}(\eta_{{\mathbf s}_2{\mathbf b}},\,{\mathbf r}_2)$ in $S^3$. Thus, as ${\mathbf s}_2\,\rightarrow\,{\mathbf b}$ during the detection process so that the angle $\eta_{{\mathbf s}_2{\mathbf b}}\rightarrow\,0$,
\begin{equation}
{\mathscr B}({\mathbf b},{\lambda})\longrightarrow\mp1, \,\;\text{with its average}\;\Bigl\langle{\mathscr B}({\mathbf b},{\lambda})\Bigr\rangle\longrightarrow 0.
\end{equation}
It is important to note that the functions (\ref{a-q}) and (\ref{b-q}) specify measurement results obtained in two separate set of experiments, independently carried out by Alice and Bob regardless of each other. The strong correlations between their results, however, are obtained when they {\it jointly} observe their results in ``coincident counts'', at a later time, within a spacelike hypersurface in spacetime, which we have assumed to be $S^3$. This requires a different set of experiments and leads to the following question: What will be the average of the product ${{\mathscr A}{\mathscr B}({\mathbf a},{\mathbf b},\lambda)}$ when the individual results ${\mathscr A}({\mathbf a},\lambda)=\pm1$ and ${\mathscr B}({\mathbf b},\lambda)=\pm1$ are observed by Alice and Bob separately but simultaneously within $S^3$? The answer to this question is given by the following special theorem.
\begin{theorem} \label{1}
The strong quantum mechanical correlations predicted by the entangled singlet state can be understood as classical, local, realistic, and deterministic correlations among the pairs of limiting scalar points ${{\mathscr A}({\bf a},\,{\lambda})=\pm1}$ and ${{\mathscr B}({\bf b},\,{\lambda})=\pm1}$ of a quaternionic 3-sphere, or $S^3$, taken as a model of the three-dimensional physical space.
\end{theorem}
I have proved this theorem in several different ways in previous publications \cite{IJTP,RSOS,IEEE-1,IEEE-2}. The proof amounts to computing the singlet correlations while preserving the geometric properties of the 3-sphere, using the following standard formula provided by Bell \cite{Bell-1964}:
\begin{equation}
{\cal E}_{\mathrm{L.R.}}({\mathbf a},\,{\mathbf b})=\int
{\mathscr A}({\mathbf a},\,\lambda)\,{\mathscr B}({\mathbf b},\,\lambda)\;p(\lambda)\,d\lambda\approx\!\lim_{\,n\,\gg\,1}\left[\frac{1}{n}\sum_{k\,=\,1}^{n}\,{\mathscr A}({\mathbf a},\,{\lambda}^k)\;{\mathscr B}({\mathbf b},\,{\lambda}^k)\right]=\,-\cos(\,\eta_{{\mathbf a}{\mathbf b}}). \label{65a}
\end{equation}
Since $\lambda$ is a fair coin with only possible values $+1$ or $-1$, the averaging over $\lambda$ can be assumed to be uniform with probability distribution $p(\lambda)=\frac{1}{n}$, where $n$ is the total number of trials. This allows us to reduce the above expectation function to the discrete version. In Section~\ref{Sec-5} below, a new proof of Theorem~\ref{1} is presented that is immune to a critique of its proofs in Christian \cite{IJTP,RSOS,IEEE-1,IEEE-2}. 

\subsection{General Theorem} \label{subsec2.2}

The above special case allows us to reproduce the predictions of simple quantum states such as the two-particle singlet state and the Hardy state \cite{Disproof}. However, for reproducing more general quantum correlations the quaternionic 3-sphere by itself is not sufficient. What is required is an algebraic representation space of the quaternionic 3-sphere defined in (\ref{nonsin}). It turns out to be a 7-sphere constructed from the eight-dimensional even sub-algebra (which I have christened ${\cal K}^{\lambda}$) of the ${2^4=16}$-dimensional Clifford algebra ${\mathrm{Cl}_{4,0}\,}$, to be discussed in the next section. The algebra ${\cal K}^{\lambda}$ happens to be a conformal counterpart of the algebra ${\mathrm{Cl}_{3,0}}$ of orthogonal directions in $\mathrm{I\!R}^3$. The general elements of ${\cal K}^{\lambda}$ can be written as a sum involving two quaternions as follows:
\begin{equation}
{\cal K}^{\lambda}:=\left\{{\mathbb Q}_z :=\,{\bf q}_r + {\bf q}_d\,\varepsilon \;\big|\;\left|\left|{\mathbb Q}_{z}\right|\right|^2={\mathbb Q}_{z}{\mathbb Q}^{\dagger}_{z}={\varrho}^2_c+{\sigma_c^2}\,\varepsilon\right\}, \label{evenspace}
\end{equation}
where $\dagger$ signifies the conjugate or reverse operation, ${\varrho}^2_c={\bf q}_{r}\,{\bf q}^{\dagger}_{r}+{\bf q}_{d}\,{\bf q}^{\dagger}_{d}=\varrho_r^2+\varrho_d^2$ and ${\sigma}_c^2={\bf q}_{r}\,{\bf q}^{\dagger}_{d}+{\bf q}_{d}\,{\bf q}^{\dagger}_{r}$ are scalar quantities, and $\varepsilon$ is a pseudoscalar that satisfies $\varepsilon^2=+1$ and $\varepsilon^{\dagger}\!=\varepsilon$. The required 7-sphere can thus be easily constructed by setting $\sigma_c^2=0$:
\begin{equation}
{\cal K}^{\lambda}\hookleftarrow S^7:=\,\left\{{\mathbb Q}_z :=\, {\bf q}_r + {\bf q}_d\,\varepsilon\;\Big|\;\sigma_c^2={\bf q}_{r}\,{\bf q}^{\dagger}_{d}+{\bf q}_{d}\,{\bf q}^{\dagger}_{r}=0\Longleftrightarrow||{\mathbb Q}_z||=\varrho_c\right\}, \label{2-7}
\end{equation}
where $\lambda=\pm1$ specifies the orientation or handedness of ${\cal K}^{\lambda}$ and $S^7$. This leads to the following generalization of Theorem~\ref{1}:
\begin{theorem} \label{2}
The quantum mechanical correlations predicted by any arbitrary quantum state can be understood as classical, local, realistic, and deterministic correlations among the pairs of limiting scalar points ${{\mathscr A}({\bf a},\,{\lambda})=\pm1}$ and ${{\mathscr B}({\bf b},\,{\lambda})=\pm1}$ of the octonion-like 7-sphere (\ref{2-7}), viewed as an algebraic representation space of the quaternionic 3-sphere considered in Theorem~\ref{1}.
\end{theorem}
Here, in analogy with (\ref{a-q}) and (\ref{b-q}), the measurement results ${{\mathscr A}({\bf a},\,{\lambda})=\pm1}$, ${{\mathscr B}({\bf b},\,{\lambda})=\pm1}$, {\it etc.}, are specified by the functions
\begin{align}
S^7\ni{\mathscr A}({\bf a}\,,\,\lambda):=&\lim_{\,\substack{{\bf s}_{r1}\,\rightarrow\;{\bf a}_r \\ \,{\bf s}_{d1}\,\rightarrow\;{\bf a}_d}}\left\{\,\pm\,{\bf D}({\bf a}_r,\,{\bf a}_d,\,0)\,{\bf L}({\bf s}_{r1},\,{\bf s}_{d1},\,0,\,\lambda)\,\right\}=\,
\begin{cases}
\,\mp\,1\;\;\;\;\;{\rm if} &\lambda\,=\,+\,1 \\
\,\pm\,1\;\;\;\;\;{\rm if} &\lambda\,=\,-\,1
\end{cases} \Bigg\}, \nonumber \\
&\;\,\text{together with}\,\;\Bigl\langle\, {\mathscr A}({\bf a}\,,\,\lambda) \,\Bigr\rangle\,=\,0\,, \label{sevendone}
\end{align}
\begin{align}
S^7\ni{\mathscr B}({\bf b}\,,\,\lambda):=&\lim_{\,\substack{{\bf s}_{r2}\,\rightarrow\;{\bf b}_r \\ \,{\bf s}_{d2}\,\rightarrow\;{\bf b}_d}}\left\{\,\mp\,{\bf D}({\bf b}_r,\,{\bf b}_d,\,0)\,{\bf L}({\bf s}_{r2},\,{\bf s}_{d2},\,0,\,\lambda)\,\right\}=\,
\begin{cases}
\,\pm\,1\;\;\;\;\;{\rm if} &\lambda\,=\,+\,1 \\
\,\mp\,1\;\;\;\;\;{\rm if} &\lambda\,=\,-\,1
\end{cases} \Bigg\}, \nonumber \\
&\;\,\text{together with}\,\;\Bigl\langle\, {\mathscr B}({\bf b}\,,\,\lambda) \,\Bigr\rangle\,=\,0\,, \label{sevenundone-1}
\end{align}
\begin{align}
S^7\ni{\mathscr C}({\bf c}\,,\,\lambda):=&\lim_{\,\substack{{\bf s}_{r3}\,\rightarrow\;{\bf c}_r \\ \,{\bf s}_{d3}\,\rightarrow\;{\bf c}_d}}\left\{\,\pm\,{\bf D}({\bf c}_r,\,{\bf c}_d,\,0)\,{\bf L}({\bf s}_{r3},\,{\bf s}_{d3},\,0,\,\lambda)\,\right\}=\,
\begin{cases}
\,\mp\,1\;\;\;\;\;{\rm if} &\lambda\,=\,+\,1 \\
\,\pm\,1\;\;\;\;\;{\rm if} &\lambda\,=\,-\,1
\end{cases} \Bigg\}, \nonumber \\
&\;\,\text{together with}\,\;\Bigl\langle\, {\mathscr C}({\bf c}\,,\,\lambda) \,\Bigr\rangle\,=\,0\,, \label{sevenundone}
\end{align}
{\it etc.}, representing arbitrary number of measurement results observed by arbitrary number of experimenters. In Christian \cite{RSOS} I have proved this theorem for arbitrary quantum states by formally computing correlations in analogy with those computed in (\ref{65a}):
\begin{equation}
{\cal E}_{\mathrm{L.R.}}({\bf a},\,{\bf b},\,{\bf c},\,{\bf d},\,{\bf e},\,{\bf f},\,\dots)\,=\!\lim_{\,n\,\gg\,1}\left[\frac{1}{n}\sum_{k\,=\,1}^{n}\,{\mathscr A}({\bf a},\,\lambda^k)\;{\mathscr B}({\bf b},\,\lambda^k)\;{\mathscr C}({\bf c},\,\lambda^k)\;{\mathscr D}({\bf d},\,\lambda^k)\;{\mathscr E}({\bf e},\,\lambda^k)\;{\mathscr F}({\bf f},\,\lambda^k)\dots\right].
\end{equation}
Moreover, as an example I have also reproduced the correlations quantum mechanically predicted by the four-particle GHZ state:
\begin{equation}
{\cal E}^{\mathrm{GHZ}}_{\mathrm{L.R.}}({\bf a},\,{\bf b},\,{\bf c},\,{\bf d})=\,\cos\theta_{\bf a}\,\cos\theta_{\bf b}\,\cos\theta_{\bf c}\,\cos\theta_{\bf d}\,-\,\sin\theta_{\bf a}\,\sin\theta_{\bf b}\,\sin\theta_{\bf c}\,\sin\theta_{\bf d}\,\cos\left(\,\phi_{\bf a}\,+\,\phi_{\bf b}\,-\,\phi_{\bf c}\,-\,\phi_{\bf d}\,\right), \label{203}
\end{equation}
where ${\theta_{\bf a}}$, ${\phi_{\bf b}}$, {\it etc.}, represent the polar and azimuthal angles, respectively, of the detector directions ${\bf a}$, ${\bf b}$, {\it etc}. This is exactly the prediction of four-particle GHZ state \cite{RSOS}. But I have derived it as a geometrical effect within the locally causal 7-sphere framework.

In fact, the Tsirel’son’s bounds on a CHSH-type correlator characterizing the strength of arbitrary quantum correlations, namely 
\begin{equation}
-2\sqrt{2}\leqslant{\cal E}({\mathbf{a}},\,{\mathbf{b}})+{\cal E}({\mathbf{a}},\,{\mathbf{b'}})+{\cal E}({\mathbf{a'}},\,{\mathbf{b}})-{\cal E}({\mathbf{a'}},\,{\mathbf{b'}})\leqslant 2\sqrt{2}\,, 
\end{equation}
also turns out to be a consequence of the quaternionic 3-sphere hypothesis \ref{hyp1}, and can be reproduced within the octonion-like 7-sphere framework constructed in Christian \cite{RSOS}. According to this hypothesis, the strong correlations observed in Nature, such as the singlet correlations \cite{Aspect}, is a compelling evidence that we live in a quaternionic 3-sphere and not in ${\mathrm{I\!R}^3}$. However, singlet correlations are usually understood as due to quantum entanglement. I have therefore proposed an experiment in Christian \cite{IJTP}, set in a macroscopic domain, so that concepts from quantum theory cannot explain the strong correlations if they are observed in that experiment. Thus the 3-sphere hypothesis is falsifiable. Considering its significance, I now turn to its mathematical foundations.

\section{Clifford-algebraic foundations of the octonion-like 7-sphere}

My goal in this section is to establish the Clifford-algebraic foundations of the 7-sphere constructed in (\ref{2-7}) by starting with the quaternionic 3-sphere defined in (\ref{nonsin}). I have explained these foundations adequately in Christian \cite{RSOS}. However, some parts of the 7-sphere framework have been challenged in Lasenby \cite{Lasenby-AACA}, making several incorrect claims. In the next section, I address those claims. But for that purpose, it will be instructive to first understand the foundations of the 7-sphere framework in some detail. 

\subsection{From even subalgebra of $\mathrm{Cl_{3,0}}$ to even subalgebra of $\mathrm{Cl_{4,0}}$}

\begin{figure}
\scalebox{0.9}{
\begin{pspicture}(-7.5,-3.0)(7.5,3.2)

\psline[linewidth=0.4mm,arrowinset=0.3,arrowsize=3pt 3,arrowlength=2]{<->}(-5.47,-2.3)(10.2,-2.3)

\psline[linewidth=0.3mm,arrowinset=0.3,arrowsize=3pt 3,arrowlength=2]{->}(-5.0,0.0)(-4.0,0.0)

\psline[linewidth=0.3mm,arrowinset=0.3,arrowsize=3pt 3,arrowlength=2]{->}(-5.0,0.0)(-5.0,1.0)

\psline[linewidth=0.2mm,arrowinset=0.3,arrowsize=3pt 3,arrowlength=2]{->}(2.35,-2.3)(8.35,-2.3)

\psline[linewidth=0.2mm,arrowinset=0.3,arrowsize=3pt 3,arrowlength=2]{->}(2.35,-2.3)(4.55,0.63)

\put(2.25,2.55){\large {${{\bf e}_{\infty}}$}}

\put(2.89,2.55){${\sim(0,2)}$}

\put(-2.9,-2.85){\large {${{\mathrm{I\!R}}^3}$}}

\put(2.3,-2.7){\large {${\rm o}$}}

\put(-4.0,0.8){\large {${{\rm I\!R}^4}$}}

\put(-5.1,1.1){\large {${\hat{x}_4}$}}

\put(-3.85,-0.1){\large {${\bf x}$}}

\put(8.45,-2.15){\large {${\bf x}$}}

\put(4.7,0.75){\large {${\vec{\phi}({\bf x})}$}}

\put(-1.99,1.25){\large {${S^3\!\sim\vec{\phi}\left({\mathrm{I\!R}}^3\right)}$}}

\put(-5.97,-2.4){{${\infty}$}}

\put(-5.97,2.25){{${\infty}$}}

\put(10.37,2.25){{${\infty}$}}

\put(10.37,-2.4){{${\infty}$}}

\psline[linewidth=0.2mm,linestyle=dashed,arrowinset=0.3,arrowsize=3pt 3,arrowlength=2]{<->}(-5.47,2.3)(10.2,2.3)

\psline[linewidth=0.2mm,linestyle=dashed](2.35,2.3)(8.35,-2.3)

\psline[linewidth=0.2mm,linestyle=dashed](2.35,2.3)(6.35,-2.3)

\psline[linewidth=0.2mm,linestyle=dashed](2.35,2.3)(4.35,-2.3)

\psline[linewidth=0.3mm,linestyle=dashed](2.35,-2.3)(2.35,2.2)

\psline[linewidth=0.1mm,dotsize=2pt 3]{*-}(2.35,2.27)(2.4,2.27)

\psline[linewidth=0.1mm,dotsize=2pt 2]{*-}(2.35,0)(2.35,0)

\put(2.45,-0.1){{${(0,1)}$}}

\psline[linewidth=0.2mm,linestyle=dashed](2.35,2.3)(0.35,-2.3)

\psline[linewidth=0.2mm,linestyle=dashed](2.35,2.3)(-1.65,-2.3)

\psline[linewidth=0.2mm,linestyle=dashed](2.35,2.3)(-3.65,-2.3)

\pscircle[linewidth=0.4mm](2.35,0){2.3}

\psline[linewidth=0.1mm,dotsize=2pt 3]{*-}(2.35,2.3)(2.4,2.3)

\psline[linewidth=0.1mm,dotsize=2.5pt 3]{*-}(2.35,-2.29)(2.4,-2.31)

\psline[linewidth=0.1mm,dotsize=2pt 3]{*-}(8.35,-2.3)(8.4,-2.3)

\psline[linewidth=0.1mm,dotsize=2pt 3]{*-}(6.35,-2.3)(6.4,-2.3)

\psline[linewidth=0.1mm,dotsize=2pt 3]{*-}(4.35,-2.3)(4.4,-2.3)

\psline[linewidth=0.1mm,dotsize=2pt 3]{*-}(0.35,-2.3)(0.34,-2.3)

\psline[linewidth=0.1mm,dotsize=2pt 3]{*-}(-1.65,-2.3)(-1.64,-2.3)

\psline[linewidth=0.1mm,dotsize=2pt 3]{*-}(-3.65,-2.3)(-3.64,-2.3)

\psline[linewidth=0.1mm,dotsize=2pt 3]{*-}(4.55,0.62)(4.55,0.64)

\psline[linewidth=0.1mm,dotsize=2pt 3]{*-}(0.15,0.61)(0.15,0.63)

\psline[linewidth=0.1mm,dotsize=2pt 3]{*-}(4.61,-0.3)(4.615,-0.3)

\psline[linewidth=0.1mm,dotsize=2pt 3]{*-}(0.09,-0.3)(0.095,-0.3)

\psline[linewidth=0.1mm,dotsize=2pt 3]{*-}(4.02,-1.55)(4.07,-1.55)

\psline[linewidth=0.1mm,dotsize=2pt 3]{*-}(0.68,-1.55)(0.7,-1.55)

\end{pspicture}}
\caption{One-point compactification of ${{\mathrm{I\!R}}^3}$ by means of a stereographic projection onto ${S^3\!\in{\rm I\!R}^4}$. After Christian \cite{RSOS}.}
\label{Fig-1}
\end{figure}
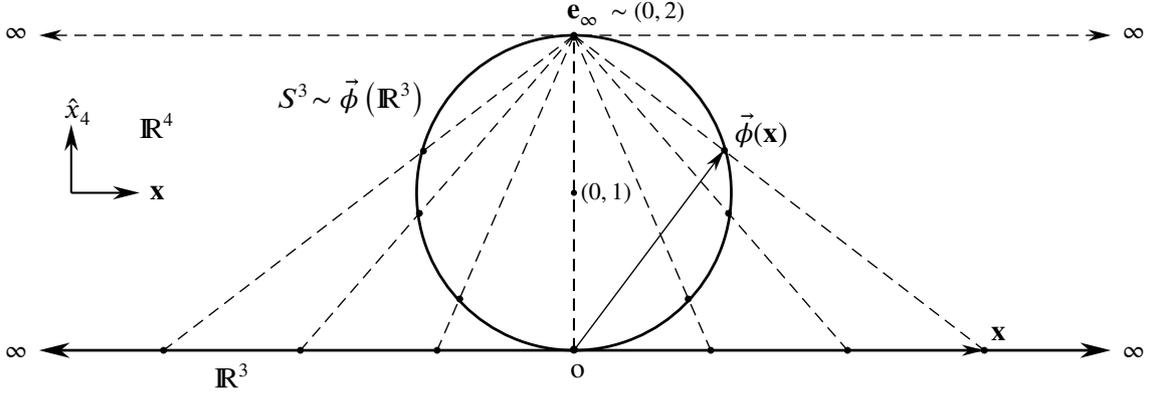

In Section~\ref{Sec2}, we saw how Bell's impossibility claim \cite{Bell-1964} can be overcome by modeling the physical space as quaternionic 3-sphere and using geometric algebra, instead of modeling it as $\mathrm{I\!R}^3$ and using vector algebra. Now it is well known that topologically $S^3$ can be constructed by adding a point to ${{\mathrm{I\!R}}^3}$ at infinity, by means of one-point or Alexandroff compactification (cf. Figure~\ref{Fig-1}\!\!\!) \cite{Nakahara}:
\begin{equation}
S^3 = \,{\mathrm {I\!R}}^3 \cup \left\{\infty\right\}. \label{onepoint}
\end{equation}
The resulting manifold is then closed and compact space, with remarkable geometrical properties \cite{IJTP,Bloch}. On the other hand, the algebraic properties of ${{\mathrm{I\!R}}^3}$ are best captured by the Clifford algebra $\mathrm{Cl_{3,0}}$ of orthogonal directions in ${{\mathrm{I\!R}}^3}$. Thus the algebra $\mathrm{Cl_{3,0}\,}$, as an eight-dimensional linear vector space spanned by graded basis, can be viewed as an algebraic representation space of ${{\mathrm{I\!R}}^3}$:
\begin{equation}
\mathrm{Cl}_{3,0}={\rm span}\!\left\{\,1,\;{\bf e}_x,\,{\bf e}_y,\,{\bf e}_z,\;{\bf e}_x{\bf e}_y,\,
{\bf e}_z{\bf e}_x,\,{\bf e}_y{\bf e}_z,\;{\bf e}_x{\bf e}_y{\bf e}_z=:I_3\,\right\}\!. \label{cl}
\end{equation}
Here ${{\bf e}^2_i=1}$ is an identity element of grade-0, ${{\bf e}_i}$ are three orthonormal vectors of grade-1, ${{\bf e}_j{\bf e}_k}$ are three orthonormal bivectors of grade-2, and ${{\bf e}_i{\bf e}_j{\bf e}_k}$ is a trivector of grade-3. Respectively, they represent points, lines, planes, and volumes in ${{\rm I\!R}^3}$. There are thus ${2^3=8}$ ways to combine the vectors ${{\bf e}_i}$ so that no two products are linearly dependent. This characterizes ${{\rm I\!R}^3}$ by the bijection
\begin{equation}
{\cal F}: {\rm I\!R}^3:={\rm span}\!\left\{\,{\bf e}_x,\,{\bf e}_y,\,{\bf e}_z\right\}\longrightarrow\, {\rm I\!R}^8:={\rm span}\!\left\{\,1,\;{\bf e}_x,\,{\bf e}_y,\,{\bf e}_z,\;{\bf e}_x{\bf e}_y,\,
{\bf e}_z{\bf e}_x,\,{\bf e}_y{\bf e}_z,\;I_3\,\right\}=\mathrm{Cl}_{3,0}. \label{arbR}
\end{equation}

The natural question for us then is: What is the analogous algebraic representation space of the closed and compact space $S^3$ defined in (\ref{nonsin}) using the even subalgebra (\ref{bi-1-m}) of $\mathrm{Cl_{3,0}}$? For the purposes of local-realistically reproducing quantum correlations more general than those predicted by the singlet state within the 3-sphere model, this is the most important question one can ask. To answer this question, let us take a closer look at the one-point compactification of $\mathrm{I\!R}^3$ depicted in Figure~\ref{Fig-1}\!\!\!. What we see in this figure is an inverse stereographic projection of $\mathrm{I\!R}^3$ onto a unit 3-sphere, ${S^3}$, by the embedding map ${\vec{\phi}:{\mathrm{I\!R}}^3\rightarrow S^3}$ given by
\begin{equation}
\vec{\phi}\left({\bf x}\in{\mathrm{I\!R}}^3\right) = \left(\frac{2}{||{\bf x}||^2+1}\right){\bf x}+\left(\frac{2\,||{\bf x}||^2}{||{\bf x}||^2+1}\right){{\hat x}_4}\,.
\end{equation}
Note that two of the dimensions of ${{\mathrm{I\!R}}^3}$ are suppressed in Figure~\ref{Fig-1}\!\!\!, and ${{\hat x}_4}$ represents the fourth dimension of the embedding space ${{\rm I\!R}^4}$. The important observation in this inverse stereographic projection is that, as an arbitrary vector ${{\bf x}\in {\mathrm{I\!R}}^3}$ from the origin approaches infinity, it is mapped to the same point ${{\bf e}_{\infty}}$ located at ${(0,2)}$, thus closing the non-compact space ${{\mathrm{I\!R}}^3}$ into a 3-sphere. Now, by shifting the origin to ${(0,1)}$, the above set of points can be inscribed also by a radial 4-vector originating from ${(0,1)}$ as
\begin{equation}
\vec{\psi}\left({\bf x}\in{\mathrm{I\!R}}^3\right) = \left(\frac{2}{||{\bf x}||^2+1}\right){\bf x}+\left(\frac{2\,||{\bf x}||^2}{||{\bf x}||^2+1}-1\right){{\hat x}_4}\,.
\end{equation}
The magnitude of this vector then confirms the unity of the radius of our conformally embedded 3-sphere within the space ${{\rm I\!R}^4}$:
\begin{equation}
1=\left|\left|\,\vec{\psi}\left({\bf x}\in{\mathrm{I\!R}}^3\right)\right|\right| = \text{radius of}\;S^3 \hookrightarrow {\rm I\!R}^4.
\end{equation}
The map ${\vec{\psi}({\bf x})}$ thus transforms the space ${{\mathrm{I\!R}}^3}$ into a unit 3-sphere within ${{\rm I\!R}^4}$, accomplishing the one-point compactification of ${{\mathrm{I\!R}}^3}$.

The one-point compactification of $S^3$ has thus led us to the embedding space ${{\mathrm{I\!R}}^4}$ spanned by the vector basis ${\left\{\,{\bf e}_x,\,{\bf e}_y,\,{\bf e}_z,\,{\bf e}_{\infty}\right\}}$, where the fourth unit vector ${\bf e}_{\infty}$ is the algebraic counterpart of the dimensionless point ${\{\infty\}}$ added to ${{\rm I\!R}^3}$ at infinity to construct $S^3$. The algebra of orthogonal directions in ${{\rm I\!R}^4}$ is then the ${2^4=16}$-dimensional Clifford algebra ${\mathrm{Cl}_{4,0}}$. Consequently, in analogy with how the even subalgebra (\ref{bi-1-m}) of $\mathrm{Cl_{3,0}}$ is used to define $S^3$ in (\ref{nonsin}), the algebraic representation space of $S^3$ can be defined using the even subalgebra of $\mathrm{Cl_{4,0}}$, which turns out to be the graded vector space we defined in Eq.~(\ref{evenspace}) of the previous section: 
\begin{equation}
{\cal K}^{\lambda}:=\,{\rm span}\!\left\{\,1,\,\lambda\,{\bf e}_x{\bf e}_y,\,\lambda\,{\bf e}_z{\bf e}_x,\,\lambda\,{\bf e}_y{\bf e}_z,\,\lambda\,{\bf e}_x{\bf e}_{\infty},\,
\lambda\,{\bf e}_y{\bf e}_{\infty},\,\lambda\,{\bf e}_z{\bf e}_{\infty},\,\lambda\,I_3{\bf e}_{\infty}\,\right\}\!. \label{BRepR}
\end{equation}
Here $\lambda=\pm$ specifies the orientation or handedness of this vector space (cf. Milnor \cite{Milnor}). For convenience, the multiplication rules satisfied by the graded basis elements of ${\cal K}^{\lambda}$ are reproduced in Table~\ref{T+1}\!\!\!. In Subsection 2.1 of Christian \cite{RSOS}, this algebra is arrived at by closing the volume form $I_3={\bf e}_x{\bf e}_y{\bf e}_z$ (which is open with the topology of $\mathrm{I\!R}^3$) with the unit vector ${\bf e}_{\infty}$, as shown in Figure~\ref{Fig-1}\!\!\!:
\begin{equation}
\mathrm{I\!R}^3\;\Longleftrightarrow\;{\bf e}_x{\bf e}_y{\bf e}_z\equiv I_3\;\longrightarrow\;I_3{\bf e}_{\infty}\equiv{\bf e}_x{\bf e}_y{\bf e}_z{\bf e}_{\infty}\;\Longleftrightarrow\;\mathrm{I\!R}^4\hookleftarrow S^3. \label{30-n}
\end{equation}
Comparing the basis elements of ${\cal K}^{\lambda}$ in (\ref{BRepR}) with those of $\mathrm{Cl_{3,0}}$ in (\ref{cl}), we see that each of the odd basis element of $\mathrm{Cl_{3,0}}$ is now closed with ${\bf e}_{\infty}$ and rendered even, as a result of the compactification implicit in (\ref{30-n}), thus making all basis elements of ${\cal K}^{\lambda}$ even.

\begin{table}[t]
\begin{center}
\SetTblrInner{rowsep=10pt}
\begin{tblr}[b]{|c|[2pt]c|c|c|c|c|c|c|c|} \hline
${*}$ &${1}$ &${\lambda\,{\bf e}_x{\bf e}_y}$ &${\lambda\,{\bf e}_z{\bf e}_x}$ &${\lambda\,{\bf e}_y{\bf e}_z}$ &${\lambda\,{\bf e}_x{\bf e}_{\infty}}$ &${\lambda\,{\bf e}_y{\bf e}_{\infty}}$ &${\lambda\,{\bf e}_z{\bf e}_{\infty}}$ &${\lambda\,I_3{\bf e}_{\infty}}$ \\ \hline[2pt]
${1}$ &${1}$ &${\lambda\,{\bf e}_x{\bf e}_y}$ &${\lambda\,{\bf e}_z{\bf e}_x}$ &${\lambda\,{\bf e}_y{\bf e}_z}$ &${\lambda\,{\bf e}_x{\bf e}_{\infty}}$ &${\lambda\,{\bf e}_y{\bf e}_{\infty}}$ &${\lambda\,{\bf e}_z{\bf e}_{\infty}}$ &${\lambda\,I_3{\bf e}_{\infty}}$ \\ \hline
${\lambda\,{\bf e}_x{\bf e}_y}$ &${\lambda\,{\bf e}_x{\bf e}_y}$ &${-1}$ &${{\bf e}_y{\bf e}_z}$ &${-{\bf e}_z{\bf e}_x}$ &${-{\bf e}_y{\bf e}_{\infty}}$ &${{\bf e}_x{\bf e}_{\infty}}$ &${I_3{\bf e}_{\infty}}$ &${-{\bf e}_z{\bf e}_{\infty}}$ \\ \hline
${\lambda\,{\bf e}_z{\bf e}_x}$ &${\lambda\,{\bf e}_z{\bf e}_x}$ &${-{\bf e}_y{\bf e}_z}$ &${-1}$ &${{\bf e}_x{\bf e}_y}$ &${{\bf e}_z{\bf e}_{\infty}}$ &${I_3{\bf e}_{\infty}}$ &${-{\bf e}_x{\bf e}_{\infty}}$ &${-{\bf e}_y{\bf e}_{\infty}}$ \\ \hline
${\lambda\,{\bf e}_y{\bf e}_z}$ &${\lambda\,{\bf e}_y{\bf e}_z}$ &${{\bf e}_z{\bf e}_x}$ &${-{\bf e}_x{\bf e}_y}$ &${-1}$ &${I_3{\bf e}_{\infty}}$ &${-{\bf e}_z{\bf e}_{\infty}}$ &${{\bf e}_y{\bf e}_{\infty}}$ &${-{\bf e}_x{\bf e}_{\infty}}$ \\ \hline
${\lambda\,{\bf e}_x{\bf e}_{\infty}}$ &${\lambda\,{\bf e}_x{\bf e}_{\infty}}$ &${{\bf e}_y{\bf e}_{\infty}}$ &${-{\bf e}_z{\bf e}_{\infty}}$ &${I_3{\bf e}_{\infty}}$ &${-1}$ &${-{\bf e}_x{\bf e}_y}$ &${{\bf e}_z{\bf e}_x}$ &${-{\bf e}_y{\bf e}_z}$ \\ \hline
${\lambda\,{\bf e}_y{\bf e}_{\infty}}$ &${\lambda\,{\bf e}_y{\bf e}_{\infty}}$ &${-{\bf e}_x{\bf e}_{\infty}}$ &${I_3{\bf e}_{\infty}}$ &${{\bf e}_z{\bf e}_{\infty}}$ &${{\bf e}_x{\bf e}_y}$ &${-1}$ &${-{\bf e}_y{\bf e}_z}$ &${-{\bf e}_z{\bf e}_x}$ \\ \hline
${\lambda\,{\bf e}_z{\bf e}_{\infty}}$ &${\lambda\,{\bf e}_z{\bf e}_{\infty}}$ &${I_3{\bf e}_{\infty}}$ &${{\bf e}_x{\bf e}_{\infty}}$ &${-{\bf e}_y{\bf e}_{\infty}}$ &${-{\bf e}_z{\bf e}_x}$ &${{\bf e}_y{\bf e}_z}$ &${-1}$ &${-{\bf e}_x{\bf e}_y}$ \\ \hline
${\lambda\,I_3{\bf e}_{\infty}}$ &${\lambda\,I_3{\bf e}_{\infty}}$ &${-{\bf e}_z{\bf e}_{\infty}}$ &${-{\bf e}_y{\bf e}_{\infty}}$ &${-{\bf e}_x{\bf e}_{\infty}}$ &${-{\bf e}_y{\bf e}_z}$ &${-{\bf e}_z{\bf e}_x}$ &${-{\bf e}_x{\bf e}_y}$ &${1}$ \\
\hline
\end{tblr}
\end{center}
\caption{Multiplication table for the geometric algebra ${\cal K}^{\lambda}$ defined in (\ref{BRepR}). Here ${I_3={\bf e}_x{\bf e}_y{\bf e}_z}$, ${{\bf e}_{\infty}^2=+1}$, and ${\lambda =\pm 1}$.}
\label{T+1}
\end{table}

\subsection{Setting up notations}

Next, we seek to close ${\cal K}^{\lambda}$ by normalizing its elements in analogy with $S^3$ defined in (\ref{nonsin}). Since $S^3$ is a closed and compact space with spherical topology, it is natural to expect its algebraic representation space to have the same topological properties. To that end, consider an arbitrary multivector in the vector space ${\cal K}^{\lambda}$ defined in (\ref{BRepR}) with the coefficients $\{q_0,\,q_1,\,q_2,\,q_3,\,q_4,\,q_5,\,q_6,\,q_7\}$:
\begin{equation}
{\mathbb Q}_z=\,q_0+q_1\,\lambda{\bf e}_x{\bf e}_y+q_2\,\lambda{\bf e}_z{\bf e}_x+q_3\,\lambda{\bf e}_y{\bf e}_z+q_4\,\lambda{\bf e}_x{\bf e}_{\infty}+q_5\,\lambda{\bf e}_y{\bf e}_{\infty}+q_6\,\lambda{\bf e}_z{\bf e}_{\infty}+q_7\,\lambda I_3{\bf e}_{\infty}\,. \label{s-non-s}
\end{equation}
For our purposes, it is convenient to write this general element of ${\cal K}^{\lambda}$ in terms of two independent quaternions defined in (\ref{nonsin}) as
\begin{equation}
{\mathbb Q}_z =\, {\bf q}_r + {\bf q}_d\,\varepsilon\,, \label{genele}
\end{equation}
so that
\begin{equation}
{\cal K}^{\lambda}=\left\{ {\mathbb Q}_z =\, {\bf q}_r + {\bf q}_d\,\varepsilon \right\}, \label{RepR}
\end{equation}
with
\begin{equation}
{\bf q}_r:=\,q_0 + q_1\,\lambda\,{\bf e}_x{\bf e}_y+q_2\,\lambda\,{\bf e}_z{\bf e}_x+q_3\,\lambda\,{\bf e}_y{\bf e}_z\,,
\end{equation}
\begin{equation}
{\bf q}_d :=\,-q_7 + q_6\,{\bf e}_x{\bf e}_y+q_5\,{\bf e}_z{\bf e}_x+q_4\,{\bf e}_y{\bf e}_z\,,
\end{equation}
and pseudoscalar ${\varepsilon:=-\lambda\,I_3{\bf e}_{\infty}}$ with its square ${\,\varepsilon^2={\bf e}^2_{\infty}=+1}$ and reverse (or conjugate) $\varepsilon^{\dagger}=\varepsilon$. The quaternions ${\bf q}_r$ and ${\bf q}_d$ can be normalized, respectively, to radii ${\varrho_r}$ and $\varrho_d$, defining two copies of a 3-sphere embedded in ${\mathrm{I\!R}}^4$ similar to that defined in (\ref{nonsin}):
\begin{equation}
S_r^3=\left\{\,{\bf q}_r=\,q_0 + q_1\,\lambda\,{\bf e}_x{\bf e}_y+q_2\,\lambda\,{\bf e}_z{\bf e}_x+q_3\,\lambda\,{\bf e}_y{\bf e}_z \;\Big|\;||{\bf q}_r|| = \sqrt{{\bf q}_r\,{\bf q}_r^{\dagger}\,}=\varrho_r \, \right\} \hookrightarrow{\mathrm{I\!R}}^4
\end{equation}
and
\begin{equation}
S_d^3=\left\{\,{\bf q}_d=-q_7 + q_6\,{\bf e}_x{\bf e}_y+q_5\,{\bf e}_z{\bf e}_x+q_4\,{\bf e}_y{\bf e}_z \;\Big|\;||{\bf q}_d|| = \sqrt{{\bf q}_d\,{\bf q}_d^{\dagger}\,}= \varrho_d \, \right\} \hookrightarrow{\mathrm{I\!R}}^4.\label{s-r}
\end{equation}
This amounts to constraining the coefficients ${q_0,q_1,q_2,q_3,q_4,q_5,q_6,}$ and $q_7$ of the vector ${\mathbb Q}_{z}$ with a scalar radius $\varrho_c$ as follows:
\begin{equation}
||{\mathbb Q}_{z}||^2={\mathbb Q}_{z}\cdot{\mathbb Q}^{\dagger}_{z}=\,
{\bf q}_{r}\,{\bf q}^{\dagger}_{r}\,+\,{\bf q}_{d}\,{\bf q}^{\dagger}_{d} \,=\, \varrho^2_r + \varrho^2_d\,=\sum_{k\,=\,0}^7 q_k^2\,=\,\varrho^2_c. \label{13a}
\end{equation}
On the other hand, it is easy to work out that the analogous quantity involved in the ``cross term'' also gives a scalar quantity:
\begin{equation}
\sigma_c^2:={\bf q}_{r}\,{\bf q}^{\dagger}_{d}\,+\,{\bf q}_{d}\,{\bf q}^{\dagger}_{r} = -2\,q_0q_7 + 2\,\lambda\,q_1q_6 + 2\,\lambda\,q_2q_5 + 2\,\lambda\,q_3q_4\,. \label{14a}
\end{equation}

\subsection{Uniformly applicable definition of norm} \label{Sec-3.3}

Unfortunately, because the scalar product ${\mathbb Q}_{z}\cdot{\mathbb Q}^{\dagger}_{z}$ used in (\ref{13a}) to evaluate norms is not a fundamental product in geometric algebra and therefore does not preserve the Clifford algebra structure of ${\cal K}^{\lambda}$, the prescription $||{\mathbb Q}_{z}||=\sqrt{{\mathbb Q}_{z}\cdot{\mathbb Q}^{\dagger}_{z}\,}$ leads to internal inconsistency when ${\mathbb Q}_{z}$ happens to be a product of two or more multivectors within ${\cal K}^{\lambda}$, such as ${\mathbb Q}_{z1}{\mathbb Q}_{z2}$ in (\ref{5}) below. As we will see in Subsection~\ref{4.3}, this internal inconsistency comes to light when the left-hand side and right-hand side of the composition law (\ref{5}) ends up requiring two incompatible products --- a combination of geometric product and scalar product on the left-hand side of (\ref{5}) but only scalar products on the right-hand side of (\ref{5}). The prescription $||{\mathbb Q}_{z}||=\sqrt{{\mathbb Q}_{z}\cdot{\mathbb Q}^{\dagger}_{z}\,}$ is also inconsistent with how the norms of the quaternions within ${\cal K}^{\lambda}$ are evaluated using geometric products: $||{\bf q}_{r}||:=\sqrt{{\bf q}_{r}\,{\bf q}^{\dagger}_{r}\,}$. We will therefore employ a definition of norm that is uniformly applicable and remains consistent when products of multivectors are involved:
\begin{definition} \label{Defi-1}
Norm $||{\mathbb Q}_{z}||$ := $\sqrt{{\mathbb Q}_{z}\,{\mathbb Q}^{\dagger}_{z}\,}$ with the coefficient of the pseudoscalar part of the geometric product ${{\mathbb Q}_{z}\,{\mathbb Q}^{\dagger}_{z}}$ set to zero.
\end{definition}
Note that this definition of norm is equivalent to the usual prescription $||{\mathbb Q}_{z}||=\sqrt{{\mathbb Q}_{z}\cdot{\mathbb Q}^{\dagger}_{z}\,}$ for any individual multivector within ${\cal K}^{\lambda}$ when geometric products are not involved. The two definitions give one and the same scalar value for the norm. To see this, let us evaluate the geometric product ${\mathbb Q}_{z}{\mathbb Q}^{\dagger}_{z}$ using (\ref{genele}), (\ref{13a}), and (\ref{14a}). In general, it turns out to take the following form:
\begin{align}
{\mathbb Q}_{z}{\mathbb Q}^{\dagger}_{z}\,&=\left({\bf q}_r + {\bf q}_d\,\varepsilon\right)\left({\bf q}_r + {\bf q}_d\,\varepsilon\right)^{\dagger}
\label{10a}\\
&=\left({\bf q}_r + {\bf q}_d\,\varepsilon\right)\left({\bf q}_r^{\dagger} + {\bf q}_d^{\dagger}\,\varepsilon\right)
\label{10ab}\\
&=\left({\bf q}_{r}\,{\bf q}^{\dagger}_{r}\,+\,{\bf q}_{d}\,{\bf q}^{\dagger}_{d}\right)+\left({\bf q}_{r}\,{\bf q}^{\dagger}_{d}\,+\,{\bf q}_{d}\,{\bf q}^{\dagger}_{r}\right)\varepsilon \label{10b} \\
&=\left(\varrho^2_r + \varrho^2_d\right) + \left(-2\,q_0q_7 + 2\,\lambda\,q_1q_6 + 2\,\lambda\,q_2q_5 + 2\,\lambda\,q_3q_4\right) \times \varepsilon \label{10c} \\
&:= \varrho^2_c+\sigma_c^2\,\varepsilon \label{11a} \\
&=\,\text{(a scalar)} \,+\, \text{(a scalar)}\times\varepsilon \label{split} \\
&=\,\text{(a scalar)} \,+\, \text{(a pseudoscalar)}. \label{23a} 
\end{align}
Thus, the geometric product ${\mathbb Q}_{z}{\mathbb Q}^{\dagger}_{z}$ resembles a hyperbolic number or split complex number instead of a scalar \cite{Dray,Sobczyk}. Definition~\ref{Defi-1} then leads to the following normalization condition for the elements ${\mathbb Q}_{z}$, giving the same value $\varrho_c$ for the norm obtained in (\ref{13a}):
\begin{equation}
\sigma_c^2={\bf q}_{r}{\bf q}^{\dagger}_{d}+{\bf q}_{d}{\bf q}^{\dagger}_{r}=-2\,q_0q_7 + 2\,\lambda\,q_1q_6 + 2\,\lambda\,q_2q_5 + 2\,\lambda\,q_3q_4=0. \label{normcon}
\end{equation}
This amounts to setting the coefficient of the pseudoscalar $\varepsilon$ to zero by assuming that the quaternions ${\bf q}_{r}$ and ${\bf q}_{d}$ are orthogonal.

\subsection{$S^7$ as an algebraic representation space of $S^3$} \label{Sec-3.4}

Equipped with a universally consistent and uniformly applicable definition of norm, we are ready to prove the following lemmas.

\begin{lemma}\label{L1}
The norms of the elements ${\mathbb Q}_{z}\in{\cal K}^{\lambda}$ defined in (\ref{10a}) by $||{\mathbb Q}_{z}||:=\sqrt{{\mathbb Q}_{z}{\mathbb Q}^{\dagger}_{z}\,}$ are positive definite, in the sense that 
\begin{equation}
||{\mathbb Q}_{z}||=0 \iff {\mathbb Q}_{z}=0\,. \label{pd}
\end{equation}
\end{lemma}
The proof of Lemma~\ref{L1} is straightforward. Since the norm of any quaternion is positive definite, the norms of the quaternions ${\bf q}_{r}$ and ${\bf q}_{d}$ are positive definite: $||{\bf q}_{r}||=0\Longleftrightarrow {\bf q}_{r}=0$ and $||{\bf q}_{d}||=0\Longleftrightarrow {\bf q}_{d}=0$. From (\ref{10b}) it is then evident that the scalar part of $||{\mathbb Q}_{z}||:=\sqrt{{\mathbb Q}_{z}{\mathbb Q}^{\dagger}_{z}}$ cannot vanish unless both ${\bf q}_{r}$ and ${\bf q}_{d}$ are vanishing. But in that case, ${\mathbb Q}_{z}=0$ according to its definition (\ref{genele}). Thus $||{\mathbb Q}_{z}||=0$ if ${\mathbb Q}_{z}=0$. Conversely, ${\mathbb Q}_{z}$ would be zero only if both ${\bf q}_{r}$ and ${\bf q}_{d}$ are zero. But in that case $||{\mathbb Q}_{z}||=0$ using (\ref{10b}).
\begin{lemma}\label{L2}
The norms of the elements ${\mathbb Q}_{z}\in{\cal K}^{\lambda}$ defined in (\ref{10a}) by $||{\mathbb Q}_{z}||:=\sqrt{{\mathbb Q}_{z}{\mathbb Q}^{\dagger}_{z}\,}$ are multiplicative, in the sense that the norm $||{\mathbb Q}_{z1}{\mathbb Q}_{z2}||$ of the product of any two elements ${\mathbb Q}_{z1}$ and ${\mathbb Q}_{z2}$ in ${\cal K}^{\lambda}$ is equal to the product of their norms $||{\mathbb Q}_{z1}||$ and $||{\mathbb Q}_{z2}||$:
\begin{equation}
||{\mathbb Q}_{z1}{\mathbb Q}_{z2}|| = ||{\mathbb Q}_{z1}||\;||{\mathbb Q}_{z2}||. \label{no-5}
\end{equation}
\end{lemma}
The proof of Lemma~\ref{L2} is given in Appendices~\ref{Appen1} below. It involves straightforward computation that demonstrates that, for any arbitrary multivectors in ${\cal K}^{\lambda}$ such as ${\mathbb Q}_{z1}$ and ${\mathbb Q}_{z2}$, both the left-hand side and the right-hand side of the composition law \cite{Eight}
\begin{equation}
||{\mathbb Q}_{z1}{\mathbb Q}_{z2}||^2\,=\;||{\mathbb Q}_{z1}||^2\;||{\mathbb Q}_{z2}||^2 \label{5}
\end{equation}
are equal to the same quantity
\begin{align}
&\left\{\left(\varrho^2_{r1}\,+\,\varrho^2_{d1}\right)\left(\varrho^2_{r2}\,+\,\varrho^2_{d2}\right) + \left({\bf q}_{r1}\,{\bf q}^{\dagger}_{d1}+{\bf q}_{d1}\,{\bf q}^{\dagger}_{r1}\right)\left({\bf q}_{r2}\,{\bf q}^{\dagger}_{d2}+{\bf q}_{d2}\,{\bf q}^{\dagger}_{r2}\right)\right\} \notag \\
&\;\;\;\;\;\;\;\;\;\;\;\;\;\;\;\;\;\;\;\;\;\;\;\;\;\;\;\;+\left\{\left( \varrho^2_{r1}\,+\,\varrho^2_{d1}\right) \left({\bf q}_{r2}\,{\bf q}^{\dagger}_{d2}+{\bf q}_{d2}\,{\bf q}^{\dagger}_{r2}\right)
+\left( \varrho^2_{r2}\,+\,\varrho^2_{d2}\right) \left({\bf q}_{r1}\,{\bf q}^{\dagger}_{d1}+{\bf q}_{d1}\,{\bf q}^{\dagger}_{r1}\right)\right\}\,\varepsilon\,.
\label{66a}
\end{align}
Recalling from (\ref{14a}) that the quantities such as ${\bf q}_{r1}\,{\bf q}^{\dagger}_{d1}+{\bf q}_{d1}\,{\bf q}^{\dagger}_{r1}$ in (\ref{66a}) are scalar quantities, we recognize that the above quantity resembles a split complex number or hyperbolic number \cite{Dray,Sobczyk} similarly to that of the product ${\mathbb Q}_{z}{\mathbb Q}^{\dagger}_{z}$ in (\ref{11a}). Consequently, as explained in Appendix \ref{Appen1}, the square-root operation can be meaningfully applied to (\ref{5}) by treating the quantity in (\ref{66a}) as a hyperbolic number $a+b\,\varepsilon$, with $a$ and $b$ being scalar numbers. This allows us to express the norm relation (\ref{5}) in square-root form as seen in (\ref{no-5}), because a product of two hyperbolic numbers, say $a_1+b_1\,\varepsilon$ and $a_2+b_2\,\varepsilon$, is also a hyperbolic number.

The norm relation (\ref{no-5}) is a general result. It holds for all multivectors ${\mathbb Q}_{z1}$ and ${\mathbb Q}_{z2}$ in ${\cal K}^{\lambda}$ without exception, in addition to ${\cal K}^{\lambda}$ being closed under multiplication. On the other hand, using the normalization condition (\ref{normcon}), the hyperbolic quantity (\ref{66a}) can be reduced to a scalar quantity by setting ${{\bf q}_{r1}\,{\bf q}^{\dagger}_{d1}+{\bf q}_{d1}\,{\bf q}^{\dagger}_{r1}=0}$ and ${{\bf q}_{r2}\,{\bf q}^{\dagger}_{d2}+{\bf q}_{d2}\,{\bf q}^{\dagger}_{r2}=0}$. This reduces the norm relation (\ref{no-5}) to
\begin{equation}
||{\mathbb Q}_{z1}\,{\mathbb Q}_{z2}|| \,=\, \sqrt{\left(\varrho^2_{r1}\,+\,\varrho^2_{d1}\right)\left(\varrho^2_{r2}\,+\,\varrho^2_{d2}\right)} \,=\, ||{\mathbb Q}_{z1}||\;||{\mathbb Q}_{z2}||\,. \label{normfinal}
\end{equation}
The advantage of this special case is that it reduces the values of the norms from hyperbolic quantities to scalar quantities as a consequence of the normalization condition (\ref{normcon}), which is rooted in the fundamental nature of the geometric product. Moreover, (\ref{normfinal}) allows us to construct the 7-sphere (\ref{2-7}) embedded in ${\mathrm{I\!R}^8}$. This is because the norm relation (\ref{normfinal}) ensures that, given two unit multivectors $X$ and $Y$ in ${\cal K}^{\lambda}$, their product $Z=XY$ would also be a unit multivector in ${\cal K}^{\lambda}$. This allows us to define $S^7$ as
\begin{equation}
{\cal K}^{\lambda}\hookleftarrow S^7:=\,\left\{\,{\mathbb Q}_z:=\,{\bf q}_r + {\bf q}_d\,\varepsilon\;\Big|\;\sigma_c^2={\bf q}_{r}\,{\bf q}^{\dagger}_{d}+{\bf q}_{d}\,{\bf q}^{\dagger}_{r}=0\Longleftrightarrow||{\mathbb Q}_z||=\sqrt{{\mathbb Q}_{z}{\mathbb Q}^{\dagger}_{z}\,}=\varrho_c\right\}. \label{sevsp}
\end{equation}
This 7-sphere is used in Theorem~\ref{2} discussed in Subsection~\ref{subsec2.2}. Its definition depends on the validity of the following lemma. 

\begin{lemma}\label{L3}
The normalization or orthogonality condition ${\sigma_c^2={\bf q}_{r}\,{\bf q}^{\dagger}_{d}+{\bf q}_{d}\,{\bf q}^{\dagger}_{r}=0}$ is preserved under multiplications of ${\mathbb Q}_{z}\in{\cal K}^{\lambda}$.
\end{lemma}
The proof of Lemma~\ref{L3} is given in Appendix~\ref{Appen2}. It proves that 7-sphere defined in (\ref{sevsp}) remains closed under multiplications of the multivectors 
${\mathbb Q}_{z}$ in ${\cal K}^{\lambda}$, despite satisfying the nontrivial normalization condition ${\sigma_c^2={\bf q}_{r}\,{\bf q}^{\dagger}_{d}+{\bf q}_{d}\,{\bf q}^{\dagger}_{r}=0}$ defined in (\ref{normcon}).
\begin{lemma} \label{L4}
The associative algebra ${{\cal K}}^{\lambda}$ defined in (\ref{RepR}) that respects the normalization condition ${\sigma_c^2={\bf q}_{r}\,{\bf q}^{\dagger}_{d}+{\bf q}_{d}\,{\bf q}^{\dagger}_{r}=0}$ defined in (\ref{normcon}) so that the norm relation (\ref{normfinal}) holds with scalar values of the norms for all multivectors ${\mathbb Q}_{z}\in{\cal K}^{\lambda}$, is a division algebra.
\end{lemma}
Given the scalar values of the norms that appear in (\ref{normfinal}), the proof of Lemma~\ref{L4} is straightforward. It is given in Appendix~\ref{Appen3} below. The 7-sphere defined in (\ref{sevsp}) is the algebraic representation space of the 3-sphere defined in (\ref{nonsin}) we sought to arrive at. Just as the algebra ${\mathrm{Cl}_{3,0}}$ is an algebraic representation space of ${\mathrm{I\!R}^3}$, the $S^7$ defined in (\ref{sevsp}) is an algebraic representation space of $S^3$.

\subsection{Comparison with Hurwitz's theorem}\label{Sec-3.5}

The claim of the original theorem by Hurwitz \cite{Hurwitz-1898,Hurwitz-1923} is that normed division algebras are possible only in 1, 2, 4, and 8 dimensions. However, the claim of the modern version of Hurwitz's theorem is much stronger \cite{Baez}: Every normed division algebra is isomorphic to either $\mathbb{R}$, $\mathbb{C}$, $\mathbb{H}$, or $\mathbb{O}$, where $\mathbb{R}$, $\mathbb{C}$, $\mathbb{H}$, and $\mathbb{O}$ stand, respectively, for the real numbers, complex numbers, quaternions, and octonions. Consequently, the results proved above appear to be in conflict with the modern claim of Hurwitz's theorem. However, the proof of Hurwitz's theorem uses inner products to compute the norms whereas I have used geometric products together with the normalization condition (\ref{normcon}), because using inner products leads to inconsistencies, as explained in Subsection~\ref{4.3} below. Consequently, the underlying coefficient algebra (\ref{split}) of ${{\cal K}^{\lambda}}$ resembles that of split complex numbers instead of real numbers. Thus the assumptions underlying the modern claim of Hurwitz's theorem are different from those underlying the above results.

\section{Response to some critiques}

Let me now turn to some critiques in Lasenby \cite{Lasenby-AACA} and Lasenby \cite{Lasenby-AGACSE-2021} of the 7-sphere framework summarized above. My point-by-point response to the critique in Lasenby \cite{Lasenby-AACA} is available online \cite{ReplyToLasenby}. However, some of the same claims were repeated in Lasenby \cite{Lasenby-AGACSE-2021}, along with some new ones, in a presentation at the AGACSE 2021 conference in honor of Professor David Hestenes. Unfortunately, these critiques have preferred to view the algebraic framework I have presented in Christian \cite{RSOS} as a ``1d up'' approach to Conformal Geometric Algebra rather than the octonion-like 7-sphere perspective I have proposed for reproducing the quantum correlations in Christian \cite{RSOS}. This has led to several misunderstandings. I address them in this section by responding to the main issues raised.

\subsection{Concerning the orientation of ${\cal K}^{\lambda}$} \label{LasA}

I have kept the orientation $\lambda$ unspecified for the vector space ${\cal K}^{\lambda}$ discussed in the previous section. It plays the role of a ``hidden variable'' or "common cause", or an initial state of the physical system within a Bell-type \cite{Bell-1964} local-realistic framework for reproducing quantum correlations. There is nothing unorthodox about this notion, as one can easily find the definition of orientation of a vector space in any good textbook on linear algebra. For example, we find the following definition of orientation in Milnor \cite{Milnor}:
\begin{definition} \label{Defi-2}
An orientation of a finite dimensional vector space ${{\cal V}_n}$ is an equivalence class of ordered basis, say ${\left\{b_1,\,\dots,\,b_n\right\}}$, which determines the same orientation of ${\,{\cal V}_n}$ as the basis ${\left\{b'_1,\,\dots,\,b'_n\right\}}$ if ${b'_i =  \omega_{ij}\, b_j}$ holds with ${{\rm det}(\omega_{ij})>0}$, and the opposite orientation of ${{\cal V}_n}$ as the basis ${\left\{b'_1,\,\dots,\,b'_n\right\}}$ if ${b'_i = \omega_{ij}\, b_j}$ holds with ${{\rm det}(\omega_{ij}) < 0}$.
\end{definition}
However, in Lasenby \cite{Lasenby-AACA} my use of an unspecified orientation $\lambda$ of the vector space ${\cal K}^{\lambda}$ is not understood and claimed to be a "problem'', without explanation. In fact, there is no problem with keeping the orientation of ${\cal K}^{\lambda}$ unspecified, because nowhere in Christian \cite{RSOS} or elsewhere have I used ${\cal K}^{+}$ and ${\cal K}^{-}$ to represent two different algebras. They both span {\it the same} algebra. Moreover, it is evident from the definition stated above that the notion of orientation is of only {\it relative} significance. If one of the vector spaces (say ${\cal K}^+$) is deemed right-handed, then the other vector space (${\cal K}^-$) would be left-handed, and vice versa. The spaces ${\cal K}^+$ and ${\cal K}^-$ are thus mathematically {\it not} identical as claimed in Lasenby  \cite{Lasenby-AACA}. This is very clearly explained in Christian \cite{RSOS} as follows:
\begin{quote}
``It is easy to verify that the bases of ${{\cal K}^+}$ and ${{\cal K}^-}$ are indeed related by an ${8\times 8}$ diagonal matrix whose determinant is ${(-1)^7 < 0}$. Consequently, ${{\cal K}^+}$ and ${{\cal K}^-}$ indeed represent right-oriented and left-oriented vector spaces, respectively, in accordance with our definition of orientation. We can therefore leave the orientation unspecified and write ${{\cal K}^{\pm}}$ as 
\begin{equation}
{\cal K}^{\lambda}=\,{\rm span}\!\left\{\,1,\,\lambda{\bf e}_x{\bf e}_y,\,\lambda{\bf e}_z{\bf e}_x,\,\lambda{\bf e}_y{\bf e}_z,\,\lambda{\bf e}_x{\bf e}_{\infty},\,\lambda{\bf e}_y{\bf e}_{\infty},\,\lambda{\bf e}_z{\bf e}_{\infty},\,\lambda I_3{\bf e}_{\infty}\,\right\}\!,\,\;\lambda^2=1\iff\lambda=\pm1.\text{''} \notag
\end{equation}
\end{quote}

\subsection{Concerning relative handedness within $S^3$} \label{Las-B}

Unfortunately, the above notion of vector space with an unspecified orientation has led to some confusion in Lasenby \cite{Lasenby-AACA} and Lasenby \cite{Lasenby-AGACSE-2021}. This has led to an instructive conceptual mistake in Lasenby \cite{Lasenby-AGACSE-2021}. Since I have already addressed it rather extensively in Questions 13, 14, and 15 in Appendix~B of Christian \cite{IEEE-2} and in Subsection~IV~D of Christian \cite{IEEE-3}, it will suffice to be brief here.

The issue raised in Lasenby \cite{Lasenby-AGACSE-2021} concerns the handedness (or orientation) $\lambda$ of the spin bivector basis (\ref{bi-1-m}) with respect to the detector bivector basis (\ref{bi-2-m}) discussed in Subsection~\ref{Sec-2.1} above. Mathematically, the relative handedness of the spins with respect to their detectors is specified by the relations (\ref{birel-m}) and (\ref{20no}). As I stressed in Subsection~\ref{Sec-2.1}, in the 3-sphere model the orientation $\lambda$ of $S^3$ has only relative meaning, in accordance with the Definition~\ref{Defi-2} of orientation stated above. For the two bivector bases $\left\{\,1,\;L_{1},\;L_{2},\;L_{3}\,\right\}$ and ${\left\{\,1,\,D_1,\,D_2,\,D_3\,\right\}}$ satisfying, respectively, the subalgebras (\ref{bi-1-m}) and (\ref{bi-2-m}), this implies the following relation:
\begin{equation}
\left(\begin{array}{c} 1 \\ L_1 \\ L_2 \\ L_3\end{array}\right)\;=\;
\begin{pmatrix}
\; 1 \;\, & \;\, 0 \;\, & \;\, 0 \;\, & \;\, 0\;\;\,\\
\, 0&\lambda&0&0\,\\
\, 0&0&\lambda&0\,\\
\, 0&0&0&\lambda
\end{pmatrix}
\;\left(\begin{array}{c} 1 \\ D_1 \\ D_2 \\ D_3 \end{array}\right).
\label{pre}
\end{equation}
Consequently, the relationship between the two bivector {\it frames} is the following:
\begin{equation}
L_1L_2L_3 = \lambda\, D_1D_2D_3 = \pm\, D_1D_2D_3. \label{corlas}
\end{equation}
This equation describes how the {\it perspectives} of the spins and the detectors are related to each other while both objectively respect the same bivector subalgebra. Moreover, physically this is not only a {\it relative} relationship but also a {\it contingent} relationship. Since the spins ${{\mathbf L}({\mathbf s})}$ and the detectors ${{\mathbf D}({\mathbf n})}$ are two entirely unrelated physical systems, Alice and Bob have no prior knowledge of the handedness of the spins until they are detected by their detectors. This makes the relationship (\ref{20no}) between the spins and detectors contingent upon the spin direction ${\mathbf s}$ being aligned to the detector direction ${\mathbf n}$ during the detection processes specified by the measurement functions (\ref{a-q}) and (\ref{b-q}). This is evident not only from these functions but also from the defining relation (\ref{20no}).

Unfortunately, in Lasenby \cite{Lasenby-AGACSE-2021} the above point of view is either overlooked or misinterpreted. For in Lasenby \cite{Lasenby-AGACSE-2021} the contingent relation $D_{i}=\lambda\,L_{i}$ from Eq.~(\ref{birel-m}) is substituted into Eq.~(\ref{bi-2-m}) that defines the bivector subalgebra respected by the detectors to obtain
\begin{equation}
L_{i}\,L_{j} \,=\,-\,\delta_{ij}\,-\lambda\sum_k\epsilon_{ijk}\,L_{k}. \label{bi-3}
\end{equation}
Similarly, by substituting the relation $L_{i}(\lambda)=\lambda\,D_{i}$ from (\ref{birel-m}) into (\ref{bi-1-m}) that defines the bivector subalgebra for spins we can obtain
\begin{equation}
D_{i}\,D_{j}\,=\,-\,\delta_{ij}\,-\lambda\sum_k\epsilon_{ijk}\,D_{k}. \label{bi-4}
\end{equation}
But such a mathematical substitution has no meaning either within the physics of the Bell-test experiments or within the quaternionic 3-sphere model \cite{IEEE-1,IEEE-2}. Indeed, comparing the above pair of equations with the defining equations (\ref{bi-1-m}) and (\ref{bi-2-m}) of spins and detectors we notice that there is an extra $\lambda$ appearing in (\ref{bi-3}) and (\ref{bi-4}) just before the summation. As a result, when $\lambda=-1$, equations (\ref{bi-3}) and (\ref{bi-4}) do not represent the bivector subalgebra of the algebra ${\mathrm{Cl}}_{(3,0)}$ and therefore do not constitute the quaternionic 3-sphere defined in (\ref{nonsin}). Thus the above mathematical step in Lasenby \cite{Lasenby-AGACSE-2021} has taken us out of the quaternionic 3-sphere model altogether. This can be verified by expressing the spin bivector basis $L_{i}$ in terms of the vector basis $\{\mathbf{e}_1',\mathbf{e}_2',\mathbf{e}_3'\}$ so that
\begin{equation}
L_1 = \mathbf{e}_2'\mathbf{e}_3', \;\;\;L_2 = \mathbf{e}_3'\mathbf{e}_1',\;\;\; \text{and}\;\;\; L_3 = \mathbf{e}_1'\mathbf{e}_2'.
\end{equation}
Straightforward textbook calculations \cite{Clifford} then give us the bivector subalgebra (\ref{bi-1-m}), not (\ref{bi-3}). Given the rules of Geometric Algebra and the properties of $\delta_{ij}$ and $\epsilon_{ijk}$, there is no way to change the sign from minus to plus appearing before the sum on the right-hand side of (\ref{bi-1-m}). Therefore, equations (\ref{bi-3}) and (\ref{bi-4}) are not correct even mathematically, let alone their physical meaninglessness.

What has been overlooked in Lasenby \cite{Lasenby-AGACSE-2021} that led us to the incorrect equations (\ref{bi-3}) and (\ref{bi-4}) with extra $\lambda$ is the fact that orientation of $S^3$ is of only {\it relative} significance between spin bivectors and detector bivectors. Since equation (\ref{bi-3}) was obtained by substituting the contingent relation $D_{i}=\lambda\,L_{i}$ from (\ref{birel-m}) into (\ref{bi-2-m}), it is meaningful {\it only with respect to} the detector bivectors defined by  (\ref{bi-2-m}). Likewise, since equation (\ref{bi-4}) was obtained by substituting the contingent relation $L_{i}=\lambda\,D_{i}$ from (\ref{birel-m}) into (\ref{bi-1-m}), it is meaningful {\it only with respect to} the spin bivectors defined by (\ref{bi-1-m}). Neither (\ref{bi-3}) nor (\ref{bi-4}) has objective meaning. The only correct interpretation of the equations (\ref{bi-1-m}) to (\ref{20no}) defining the two sets of bases within $S^3$ is how I have interpreted them in Christian \cite{IEEE-1} and Christian \cite{IEEE-2}. As demonstrated in Christian \cite{IEEE-2}, the mutual perspectives of the spin and detector bivectors are related as follows:
\begin{equation}
{\mathbf L}({\mathbf a},\,{\lambda}=+1)\;{\mathbf L}({\mathbf b},\,{\lambda}=+1)={\mathbf D}({\mathbf a})\;{\mathbf D}({\mathbf b}),
\end{equation}
and
\begin{equation}
{\mathbf L}({\mathbf a},\,{\lambda}=-1)\;{\mathbf L}({\mathbf b},\,{\lambda}=-1)={\mathbf D}({\mathbf b})\;{\mathbf D}({\mathbf a}).
\end{equation}
In other words, the order of the product of two spin bivectors and the order of the product of two detector bivectors is mutually opposite. This is what the physics of the EPR-Bohm experiments within the quaternionic 3-sphere model dictates. Accordingly, when the hidden variable or initial state ${\lambda}$ happens to be equal to ${+1}$, ${{\mathbf L}({\mathbf a},{\lambda}){\mathbf L}({\mathbf b},{\lambda})={\mathbf D}({\mathbf a}){\mathbf D}({\mathbf b})}$, and when the hidden variable or initial state ${\lambda}$ happens to be equal to ${-1}$, 
${{\mathbf L}({\mathbf a},{\lambda}){\mathbf L}({\mathbf b},{\lambda})={\mathbf D}({\mathbf b}){\mathbf D}({\mathbf a})}$. As  result, the expectation value (\ref{65a}) reduces at once to
\begin{align}
{\cal E}_{\mathrm{L.R.}}({\mathbf a},{\mathbf b})&=\frac{1}{2}\{\,{\mathbf D}({\mathbf a}){\mathbf D}({\mathbf b})\}+\frac{1}{2}\{\,{\mathbf D}({\mathbf b}){\mathbf D}({\mathbf a})\} \\
&=\frac{1}{2}(I\cdot{\mathbf a})(I\cdot{\mathbf b})+\frac{1}{2}(I\cdot{\mathbf b})(I\cdot{\mathbf a}) \\ &=-\,\frac{1}{2}\left\{{\mathbf a}{\mathbf b}+{\mathbf b}{\mathbf a}\right\}=-\,{\mathbf a}\cdot{\mathbf b}\,, \label{not63}
\end{align}
because the orientation ${\lambda}$ of ${S^3}$ is a fair coin with 50/50 chance. Here the last equality follows from the definition of inner product.

In Section~\ref{Sec-5} below I have presented an alternative derivation of the singlet correlations (\ref{65a}) within the quaternionic 3-sphere model that is immune to the critique in Lasenby \cite{Lasenby-AGACSE-2021}, because it is based on a fixed orientation, $\lambda\equiv+1$, between spins and detectors.

\subsection{Concerning the validity of composition law }\label{4.3}

In Lasenby \cite{Lasenby-AACA} and Lasenby \cite{Lasenby-AGACSE-2021} another incorrect claim is made. It concerns the proof in Christian \cite{RSOS} that the norm relation (\ref{no-5}) or (\ref{normfinal}) discussed above holds for {\it all} elements $X$ and $Y$ of ${\cal K}^{\lambda}$. As we saw in Subsection~\ref{Sec-3.4}, the proof of (\ref{normfinal}) is quite straightforward. It can be found in Subsection~2.5 of Christian \cite{RSOS}, and, with greater detail, in Appendix~\ref{Appen1} below. However, in Lasenby \cite{Lasenby-AACA} and Lasenby \cite{Lasenby-AGACSE-2021} it is claimed that the norm relation (\ref{normfinal}) is not correct, and an attempt is made to disprove it with a counterexample.

Unfortunately, the alleged counterexample is based on an internal inconsistency. Two incompatible product rules are used on the two sides of the norm relation (\ref{no-5}) to demonstrate contradiction. Using the pseudoscalar $\varepsilon$ involved in the algebra ${\cal K}^{\lambda}$ that satisfies the properties $\varepsilon^{\dagger}=\varepsilon$ and $\varepsilon^2=1$, the following two multivectors within ${\cal K}^{\lambda}$ are considered to arrive at a contradiction:
\begin{equation}
X=\frac{1}{\sqrt{2}}(1+\varepsilon)\;\;\;\;\text{and}\;\;\;\;Y=\frac{1}{\sqrt{2}}(1-\varepsilon). \label{20a}
\end{equation}
But, to begin with, no such two-dimensional objects play any role whatsoever in the 7-sphere framework presented in Christian \cite{RSOS}. Thus, even if the above multivectors are mathematically admissible, they have no relevance for the physical framework presented in Christian \cite{RSOS}. Next, scalar products ${X\cdot X^{\dagger}}=||X||^2$
and ${Y\cdot Y^{\dagger}}=||Y||^2$ are used to evaluate the norms $||X||$ and $||Y||$, giving
\begin{equation}
||{X}||=\left|\left|\frac{1}{\sqrt{2}}(1+\varepsilon)\right|\right|=\sqrt{\frac{1}{2}(1+\varepsilon)\cdot(1+\varepsilon)^{\dagger}} =\sqrt{\frac{1}{2}(1+1)}=1 \label{49-n}
\end{equation}
and
\begin{equation}
||{Y}||=\left|\left|\frac{1}{\sqrt{2}}(1-\varepsilon)\right|\right|=\sqrt{\frac{1}{2}(1-\varepsilon)\cdot(1-\varepsilon)^{\dagger}} =\sqrt{\frac{1}{2}(1+1)}=1. \label{50-n}
\end{equation}
Substituting for these values into the right-hand side of (\ref{no-5}) therefore gives $||X||\,||Y||=1$. On the other hand, to evaluate the left-hand side of (\ref{no-5}), a geometric product between the two multivectors $X$ and $Y$ is used in Lasenby \cite{Lasenby-AACA} and Lasenby \cite{Lasenby-AGACSE-2021} to obtain
\begin{equation}
||XY||=\left|\left|\frac{1}{2}(1+\varepsilon)(1-\varepsilon)\right|\right| =\frac{1}{2}\left|\left|(1-\varepsilon^2)\right|\right|=||(1-1)||=||\,0\,||=0. \label{51-n}
\end{equation}
Indeed, not using geometric product on the left-hand side of (\ref{no-5}) is not an option because that would not result in $XY\in{\cal K}^{\lambda}$. But since two incompatible product rules are used on the two sides of (\ref{no-5}), it is not surprising that desired contradiction is achieved: 
\begin{equation}
0=||{X}{Y}||\not=||{X}||\,||{Y}||=1.
\end{equation}
Since not using the geometric product for $XY$ on the left-hand side of (\ref{no-5}) is not a meaningful operation within ${\cal K}^{\lambda}$, the only way to avoid the inconsistency in what is considered in Lasenby \cite{Lasenby-AACA} and Lasenby \cite{Lasenby-AGACSE-2021} is by using geometric products throughout the above calculation, as discussed in Lemma~\ref{L2} of Subsection~\ref{Sec-3.4}. Then the correct calculation of the norms $||X||$ and $||Y||$ gives
\begin{equation}
||X||=\left|\left|\frac{1}{\sqrt{2}}(1+\varepsilon)\right|\right|=\sqrt{\frac{1}{2}(1+\varepsilon)(1+\varepsilon)^{\dagger}\,}=\sqrt{\frac{1}{2}(1+\varepsilon+\varepsilon+\varepsilon^2)\,}=\sqrt{1+\varepsilon\,}
\end{equation}
and
\begin{equation}
||Y||=\left|\left|\frac{1}{\sqrt{2}}(1-\varepsilon)\right|\right|=\sqrt{\frac{1}{2}(1-\varepsilon)(1-\varepsilon)^{\dagger}\,}=\sqrt{\frac{1}{2}(1-\varepsilon-\varepsilon+\varepsilon^2)\,}=\sqrt{1-\varepsilon\,},
\end{equation}
which now gives 0 instead of 1 for the right-hand side of (\ref{no-5}): 
\begin{equation}
||X||\,||Y||=\left(\sqrt{1+\varepsilon}\,\right)\left(\sqrt{1-\varepsilon}\,\right)=\sqrt{(1+\varepsilon)(1-\varepsilon)} =\sqrt{1-\varepsilon^2}=0. \label{53-n}
\end{equation}
Comparing (\ref{51-n}) and (\ref{53-n}) we see that now both the left-hand side $||XY||$ and the right-hand side $||X||\,||Y||$ of (\ref{no-5}) have the same value (namely, 0), and consequently the norm relation $||XY||=||X||\,||Y||$ is satisfied. As a result, the counterexample to (\ref{no-5}) alleged in Lasenby \cite{Lasenby-AACA} and Lasenby \cite{Lasenby-AGACSE-2021} fails. This is not surprising because, as discussed in Subsection~\ref{Sec-3.4} above and proved explicitly in Appendix~\ref{Appen1} below in full generality, the norm relation (\ref{no-5}) holds for any arbitrary multivectors $X$ and $Y$ in ${\cal K}^{\lambda}$.

It is instructive here to understand how the normalization condition ${\bf q}_{r}{\bf q}^{\dagger}_{d}+{\bf q}_{d}{\bf q}^{\dagger}_{r}=0$ is satisfied in this case for the product $XY$. To that end, using (\ref{genele}) and (\ref{20a}), let us compare the coefficients of the two multivectors $X$ and $Y$ with those defined in (\ref{genele}):
\begin{equation}
X=\frac{1}{\sqrt{2}}(1+\varepsilon) = {\mathbb Q}_{z1} =\, {\bf q}_{r1} + {\bf q}_{d1}\,\varepsilon
\end{equation}
and
\begin{equation}
Y=\frac{1}{\sqrt{2}}(1-\varepsilon) = {\mathbb Q}_{z2} =\, {\bf q}_{r2} + {\bf q}_{d2}\,\varepsilon\,. \label{77}
\end{equation}
Thus, the coefficients for $X$ and $Y$ chosen in (\ref{20a}) are the following: ${\bf q}_{r1}=\frac{1}{\sqrt{2}}$, ${\bf q}_{d1}=\frac{1}{\sqrt{2}}$, ${\bf q}_{r2}=\frac{1}{\sqrt{2}}$, and ${\bf q}_{d2}=\frac{-1}{\sqrt{2}}$. As a result, we have: ${\varrho^2_{r1}={\bf q}_{r1}{\bf q}^{\dagger}_{r1}=\frac{1}{2}}$, ${\varrho^2_{d1}={\bf q}_{d1}{\bf q}^{\dagger}_{d1}=\frac{1}{2}}$, ${\varrho^2_{r2}={\bf q}_{r2}{\bf q}^{\dagger}_{r2}=\frac{1}{2}}$, ${\varrho^2_{d2}={\bf q}_{d2}{\bf q}^{\dagger}_{d2}=\frac{1}{2}}$, ${{\bf q}_{r1}{\bf q}^{\dagger}_{d1}=\frac{1}{2}}$, ${{\bf q}_{d1}{\bf q}^{\dagger}_{r1}=\frac{1}{2}}$, ${{\bf q}_{r2}{\bf q}^{\dagger}_{d2}=\frac{-1}{2}}$, and ${{\bf q}_{d2}{\bf q}^{\dagger}_{r2}=\frac{-1}{2}}$. Substituting these into the quantity (\ref{66a}) that is equal to both left-hand and right-hand sides of (\ref{5}) reduces (\ref{66a}) to
\begin{equation}
\left\{\left(\frac{1}{2}+\frac{1}{2}\right)\left(\frac{1}{2}+\frac{1}{2}\right) + \left(\frac{1}{2}+\frac{1}{2}\right)\left(-\frac{1}{2}-\frac{1}{2}\right)\right\} + \left\{\left(\frac{1}{2}+\frac{1}{2}\right)\left(-\frac{1}{2}-\frac{1}{2}\right) + \left(\frac{1}{2}+\frac{1}{2}\right)\left(\frac{1}{2}+\frac{1}{2}\right)\right\}\varepsilon = \{0\} + \{0\}\,\varepsilon = 0. \label{yup}
\end{equation}
This again confirms the results (\ref{51-n}) and (\ref{53-n}) for the two sides of the norm relation (\ref{5}). But more importantly, we see from (\ref{yup}) that the coefficient of the pseudoscalar $\varepsilon$ vanishes, and consequently the normalization condition (\ref{normcon}) is automatically satisfied.

\subsection{Concerning a subsequent incorrect claim}

In Lasenby \cite{Lasenby-AACA} a further incorrect claim is made that if $X$ and $Y$ "are unit vectors, it does not follow that $Z = XY$ is also a unit vector, despite what Christian says in his Eq. (2.41) [of Christian~\cite{RSOS}].'' But now it is easy to appreciate that this claim is not correct because it depends on the validity of the counterexample alleged in Lasenby \cite{Lasenby-AACA}, which I have just shown to be invalid. Consequently, contrary to the claim in Lasenby \cite{Lasenby-AACA}, Eq. (2.41) of Christian \cite{RSOS} is quite correct. Since the norm relation (\ref{normfinal}) holds for any multivectors $X$ and $Y$ in ${\cal K}^{\lambda}$ that have scalar values for their norms, if $X$ and $Y$ happen to be unit vectors so that $||X||=1$ and $||Y||=1$, then (\ref{normfinal}) necessitates that their product $Z = XY$ will also be a unit vector: $||Z||=||XY||=||X||\,||Y||=1$.

\subsection{Concerning the reduction in dimensions}

In Lasenby \cite{Lasenby-AACA} it is claimed that ``... while the scalar part of $X\tilde{X}$ sets up a nice match with $S^7$, the extra constraint from $\langle{X\tilde{X}}\rangle_4=0$ reduces the overall dimension down to 6, ...'' However, this claim is also incorrect. Here what is meant by $\langle{X\tilde{X}}\rangle_4=0$ is the vanishing of the coefficient of the pseudoscalar as in the normalization condition (\ref{normcon}). But it is easy to see from this condition that it only reduces the eight dimensions of ${\cal K}^{\lambda}$ to the seven dimensions of $S^7$, precisely as one would expect from the relationship between the embedding space ${\cal K}^{\lambda}\sim\mathrm{I\!R}^8$ and the embedded space $S^7$. To appreciate this explicitly, let us express (\ref{normcon}) as follows:
\begin{equation}
{q}_0=\frac{\lambda}{{q}_7}({q}_1{q}_6 + {q}_2{q}_5 + {q}_3{q}_4). \label{75itis}
\end{equation}
This shows that the condition (\ref{normcon}) can be used to eliminate only {\it one} of the eight coefficients of ${\mathbb Q}_z\in{\cal K}^{\lambda}$ defined in (\ref{s-non-s}), say $q_0\,$, by writing it in terms of the remaining seven coefficients of ${\mathbb Q}_z$, as in the above equation. But that still leaves {\it seven} numbers that would be necessary to specify a point in ${\cal K}^{\lambda}$ located at the radius $\varrho_c$ from the origin, not six numbers as claimed in Lasenby \cite{Lasenby-AACA}. Moreover, this reduction of one dimension while transitioning from the eight dimensions of ${\cal K}^{\lambda}\sim\mathrm{I\!R}^8$ to the seven dimensions of $S^7$ as a result of the normalization condition (\ref{normcon}) or (\ref{75itis}) is perfectly consistent with the one less dimension of the spheres $S^0$, $S^1$, $S^3$, and $S^7$ compared to that of the corresponding normed division algebras $\mathbb{R}$, $\mathbb{C}$, $\mathbb{H}$, and $\mathbb{O}$ mentioned in Subsection~\ref{Sec-3.5}. 

\section{Alternative Derivation of the singlet correlations within $S^3$} \label{Sec-5}

Let me now present an alternative proof of Theorem~\ref{1} stated in Subsection~\ref{Sec-2.1} that does not depend on two different orientations of $S^3$ for the spin bivectors and detector bivectors. In other words, in what follows the orientation of $S^3$ is fixed to the value $\lambda=+1$, and kept fixed throughout the experiment for both the spin bivectors and detector bivectors. This makes the derivation of the singlet correlations immune to the kind of misinterpretation of $\lambda=\pm1$ in Lasenby \cite{Lasenby-AGACSE-2021} we discussed in Subsection~\ref{Las-B} above.

Since the handedness of spin bivectors and detector bivectors is now fixed to be $\lambda=+1$, they both satisfy the right-handed bivector subalgebra \cite{Clifford}. However, because the spins and detectors are different physical systems, unrelated to each other until the acts of measurements, we will continue to represent them using different notations, $L_i$ for spin basis and $D_i$ for detector basis: 
\begin{equation}
L_{i}\,L_{j} \,=\,-\,\delta_{ij}\,-\sum_k\epsilon_{ijk}\,L_{k}
\end{equation}
and
\begin{equation}
D_{i}\,D_{j}\,=\,-\,\delta_{ij}\,-\sum_k\epsilon_{ijk}\,D_{k}.
\end{equation}
Since the binary freedom of orientation $\lambda$ is no longer available, instead of (\ref{20no}) let us introduce the following two sign-functions:
\begin{equation}
\mu_1 =\text{sign}(\mathbf{a}\cdot\mathbf{s}^i_1)=\pm1
\;\;\;\text{and}\;\;\;
\mu_2 =\text{sign}(\mathbf{s}^i_2\cdot\mathbf{b})=\pm1, \label{79-nom}
\end{equation}
where the subscripts 1 and 2 on $\mu$ and ${\mathbf s}$ refer to the remote observation stations of the experimenters Alice and Bob (cf. Figure~\ref{Fig-2}\!\!\!), and the superscript $i$ on ${\mathbf s}$ indicates its initial direction at the source with respect to the eventually chosen detector directions. Moreover, since $\lambda$ is no longer available as a hidden variable, let the spin direction ${\mathbf s}={\mathbf s}_1={\mathbf s}_2$ now act as a hidden variable, just as in Bell's original local model \cite{Bell-1964,Peres}. Then the function ${\mu_1=\mathrm{sign}({\mathbf a}\cdot{\mathbf s}^i_1)}$ can be understood as follows. If, initially ({\it i.e.}, before the detection process defined by the measurement functions to be specified below), the two unit vectors ${\mathbf a}$ and ${\mathbf{s}^i_1}$ happen to be pointing through the same hemisphere of $S^2\hookrightarrow\mathrm{I\!R}^3$ centered at the origin of ${\mathbf{s}^i_1}$, then $\mu_1=\mathrm{sign}({\mathbf a}\cdot{\mathbf s}^i_1)$ will be equal to $+1$, and if the two unit vectors ${\mathbf a}$ and ${\mathbf{s}^i_1}$ happen to be pointing through the opposing hemispheres of $S^2$ centered at the origin of ${\mathbf{s}^i_1}$, then $\mu_1=\mathrm{sign}({\mathbf a}\cdot{\mathbf s}^i_1)$ will be equal to $-1$, provided that ${{\mathbf a}\cdot{\mathbf s}^i_1\not=0}$. If ${{\mathbf a}\cdot{\mathbf s}^i_1}$ happens to be zero, then $\mu_1=\mathrm{sign}({\mathbf a}\cdot{\mathbf s}^i_1)$ will be assumed to be equal to the sign of the first nonzero component from the set ${\{a_x,\,a_y,\,a_z\}}$. And likewise for the function ${\mu_2=\mathrm{sign}({\mathbf s}^i_2\cdot{\mathbf b})}$.

Next, Theorem~\ref{1} says that to explain the singlet correlations local-realistically we should consider correlations among pairs of limiting scalar points of the 3-sphere defined in (\ref{nonsin}), which is a set of unit quaternions. We therefore use products of the spin bivectors ${\mathbf L}({\mathbf s}_1)=I_3{\mathbf s}_1$ and ${\mathbf L}({\mathbf s}_2)=I_3{\mathbf s}_2$ with detector bivectors ${\mathbf D}({\mathbf a})=I_3{\mathbf a}$ and ${\mathbf D}({\mathbf b})=I_3{\mathbf b}$ using the trivector $I_3={\bf e}_x{\bf e}_y{\bf e}_z$ to represent the physical interactions that take place during the detection processes at the two ends of a Bell-test experiment. Following Bell \cite{Bell-1964}, these can be expressed as measurements functions ${\mathscr A}({\mathbf a},{\mathbf s}_1)$ and ${\mathscr B}({\mathbf b},{\mathbf s}_2)$ for the experimenters Alice and Bob:
\begin{align}
S^3\ni{\mathscr A}({\mathbf a},{\mathbf s}_1)\,&=\lim_{{\mathbf s}_1\,\rightarrow\,\mu_1{\mathbf a}}\left\{-\,{\mathbf D}({\mathbf a})\,{\mathbf L}({\mathbf s}_1)\right\} \label{79-nmn} \\
&=\lim_{{\mathbf s}_1\,\rightarrow\,\mu_1{\mathbf a}}\left\{-(I_3{\mathbf a})(I_3{\mathbf s}_1)\right\} \\
&=\lim_{{\mathbf s}_1\,\rightarrow\,\mu_1{\mathbf a}}\left\{{\mathbf a}\cdot{\mathbf s}_1+I_3({\mathbf a}\times{\mathbf s}_1)\right\} \\
&=\lim_{{\mathbf s}_1\,\rightarrow\,\mu_1{\mathbf a}}\left\{\cos(\eta_{{\mathbf a}{\mathbf s}_1})+(I_3{\mathbf r}_{1})\sin(\eta_{{\mathbf a}{\mathbf s}_1})\right\} \\
&=\lim_{{\mathbf s}_1\,\rightarrow\,\mu_1{\mathbf a}}\left\{\,+\,{\mathbf q}(\eta_{{\mathbf a}{\mathbf s}_1},\,{\mathbf r}_1)\right\}, \label{84-nom}
\end{align}
where $\eta_{{\mathbf a}{\mathbf s}_1}$ is the angle between the spin direction ${\mathbf s}_1$ and the detector direction ${\mathbf a}$ chosen by Alice (cf. Figure~\ref{Fig-2}\!\!\!), and ${\mathbf r}_1$ is the rotation axis of the quaternion ${\mathbf q}(\eta_{{\mathbf a}{\mathbf s}_1},\,{\mathbf r}_1)$ as in (\ref{rot1}). Thus, given the definition (\ref{79-nom}) of $\mu_1$, the spin direction ${\mathbf s}_1$ will tend to $+{\mathbf a}$ if initially the two unit vectors ${\mathbf a}$ and ${\mathbf{s}_1}$ happen to be pointing through the same hemisphere of $S^2$ centered at the origin of ${\mathbf{s}_1}$, and otherwise the spin direction ${\mathbf s}_1$ will tend to $-{\mathbf a}$. Consequently, as ${\mathbf s}_1\rightarrow\mu_1{\mathbf a}=\pm{\mathbf a}$ during the detection process by Alice so that the angle $\eta_{{\mathbf a}{\mathbf s}_1}\!\rightarrow0$ or $\pi$, the binary value of the observed result by Alice is obtained because $\cos(\eta_{{\mathbf a}{\mathbf s}_1})\rightarrow \pm1$ and $\sin(\eta_{{\mathbf a}{\mathbf s}_1})\rightarrow 0$: 
\begin{equation}
{\mathscr A}({\mathbf a},{\mathbf s}_1)\longrightarrow+\mu_1=\pm1. \label{85-nom}
\end{equation}
Similarly, we can define the measurement function for Bob as
\begin{align}
S^3\ni{\mathscr B}({\mathbf b},{\mathbf s}_2)\,&=\lim_{{\mathbf s}_2\,\rightarrow\,\mu_2{\mathbf b}}\left\{+\,{\mathbf L}({\mathbf s}_2)\,{\mathbf D}({\mathbf b})\right\} \label{85-nmn} \\
&=\lim_{{\mathbf s}_2\,\rightarrow\,\mu_2{\mathbf b}}\left\{+(I_3{\mathbf s}_2)(I_3{\mathbf b})\right\} \\
&=\lim_{{\mathbf s}_2\,\rightarrow\,\mu_2{\mathbf b}}\left\{-\,{\mathbf s}_2\cdot{\mathbf b}-I_3({\mathbf s}_2\times{\mathbf b})\right\} \\
&=\lim_{{\mathbf s}_2\,\rightarrow\,\mu_2{\mathbf b}}\left\{-\cos(\eta_{{\mathbf s}_2{\mathbf b}})-(I_3{\mathbf r}_{2})\sin(\eta_{{\mathbf s}_2{\mathbf b}})\right\} \\
&=\lim_{{\mathbf s}_2\,\rightarrow\,\mu_2{\mathbf b}}\left\{\,-\,{\mathbf q}(\eta_{{\mathbf s}_2{\mathbf b}},\,{\mathbf r}_2)\right\}, \label{90-nom}
\end{align}
where $\eta_{{\mathbf s}_2{\mathbf b}}$ is the angle between the spin direction ${\mathbf s}_2$ and the detector direction ${\mathbf b}$ chosen by Bob (cf. Figure~\ref{Fig-2}\!\!\!), and ${\mathbf r}_2$ is the rotation axis of the quaternion ${\mathbf q}(\eta_{{\mathbf s}_2{\mathbf b}},\,{\mathbf r}_2)$ as in (\ref{rot2}). Consequently, as ${\mathbf s}_2\rightarrow\mu_2{\mathbf b}=\pm{\mathbf b}$ during the detection process by Bob so that the angle $\eta_{{\mathbf s}_2{\mathbf b}}\!\rightarrow0$ or $\pi$, the binary value of the observed result by Bob is obtained because $\cos(\eta_{{\mathbf s}_2{\mathbf b}})\rightarrow \pm1$ and $\sin(\eta_{{\mathbf s}_2{\mathbf b}})\rightarrow 0$:
\begin{equation}
{\mathscr B}({\mathbf b},{\mathbf s}_2)\longrightarrow-\mu_2=\mp1. \label{91-nom}
\end{equation}
Thus $\mu_1$ and $\mu_2$ play the same role as the orientation $\lambda$ of $S^3$ played in the functions (\ref{a-q}) and (\ref{b-q}). It follows from (\ref{85-nom}) and (\ref{91-nom}) that, in general, for the choices ${\mathbf a}\not={\mathbf b}$, the product of the results observed by Alice and Bob will fluctuate between $-1$ and $+1$:
\begin{equation}
{\mathscr A}({\mathbf a},{\mathbf s}_1)\,{\mathscr B}({\mathbf b},{\mathbf s}_2)=\left[\lim_{{\mathbf s}_1\,\rightarrow\,\mu_1{\mathbf a}}\left\{-\,{\mathbf D}({\mathbf a})\,{\mathbf L}({\mathbf s}_1)\right\}\right]\left[\lim_{{\mathbf s}_2\,\rightarrow\,\mu_2{\mathbf b}}\left\{+\,{\mathbf L}({\mathbf s}_2)\,{\mathbf D}({\mathbf b})\right\}\right] = -\mu_1\mu_2=-\,(\pm1)(\pm1)=\mp1. \label{92-nom}
\end{equation}
On the other hand, for the choice ${\mathbf b}={\mathbf a}$, we will have ${\mu}_2={\mu}_1$, provided ${\mathbf s}_2=\,{\mathbf s}_1$, and perfect anti-correlation will be observed:
\begin{equation}
{\mathscr A}({\mathbf a},{\mathbf s}_1)\,{\mathscr B}({\mathbf a},{\mathbf s}_2)=\left[\lim_{{\mathbf s}_1\,\rightarrow\,\mu_1{\mathbf a}}\left\{-\,{\mathbf D}({\mathbf a})\,{\mathbf L}({\mathbf s}_1)\right\}\right]\left[\lim_{{\mathbf s}_2\,\rightarrow\,\mu_1{\mathbf a}}\left\{+\,{\mathbf L}({\mathbf s}_2)\,{\mathbf D}({\mathbf a})\right\}\right] =-(\mu_1)^2=-1.
\end{equation}

Note that the conservation of zero spin angular momentum requires the equality ${\mathbf s}_1=\,{\mathbf s}_2$ to hold during the free evolution of the spins, but not necessarily during their detection processes. Consequently, while the physical interactions are taking place during the detection processes at the two ends of the experiment, ${\mathbf s}_2$ can tend to $\mu_2{\mathbf b}$ jointly as ${\mathbf s}_1$ tends to $\mu_1{\mathbf a}$, despite the fact that ${\mathbf s}_1=\,{\mathbf s}_2$ must hold during the free evolution of the spins. If now the direction ${\mathbf s}={\mathbf s}_1={\mathbf s}_2$ of the spins originating at the source is assumed to be uniformly distributed over $S^2$ as in Bell's local model \cite{Bell-1964,Peres}, specified by a normalized probability measure $p({\mathbf s})$, 
\begin{equation}
\int_{S^2}p({\mathbf s})\;d{\mathbf s}=1,
\end{equation}
then the results (\ref{85-nom}) observed by Alice will be equal to $+1$ or $-1$ with 50/50 chance, and likewise for the results (\ref{91-nom}) observed by Bob. Consequently, in analogy with the quantum mechanical predictions, the expectation values of their results will vanish:
\begin{equation}
{\cal E}_{\mathrm{L.R.}}(\mathbf{a})=\int_{S^2} \mathscr{A}({\mathbf a},\mathbf{s}_1)\,p({\mathbf s})\,d{\mathbf s} = 0\;\;\;\;\text{and}\;\;\;\;{\cal E}_{\mathrm{L.R.}}(\mathbf{b})=\int_{S^2}\mathscr{B}({\mathbf b},\mathbf{s}_2)\,p({\mathbf s})\,d{\mathbf s}=0,
\end{equation}
where the subscripts 1 and 2 on the spin directions are retained for clarity even though they are the same direction (cf. Figure~\ref{Fig-2}\!\!\!).

The question now is: What will be the correlations within $S^3$ between the results ${\mathscr A}$ and ${\mathscr B}$ observed {\it jointly} but independently by Alice and Bob, in coincident counts, at a space-like distance from each other? To work out the correlations, we begin with
\begin{equation}
{\cal E}_{\mathrm{L.R.}}({\mathbf a},{\mathbf b})=\int_{S^2}{\mathscr A}({\mathbf a},{\mathbf s}_1)\;{\mathscr B}({\mathbf b},{\mathbf s}_2)\,p({\mathbf s})\,d{\mathbf s} =\int_{S^2}\left[\lim_{{\mathbf s}_1\,\rightarrow\,\mu_1{\mathbf a}}\left\{+\,{\mathbf q}(\eta_{{\mathbf a}{\mathbf s}_1},\,{\mathbf r}_1)\right\}\right]\left[\lim_{{\mathbf s}_2\,\rightarrow\,\mu_2{\mathbf b}}\left\{-\,{\mathbf q}(\eta_{{\mathbf s}_2{\mathbf b}},\,{\mathbf r}_2)\right\}\right]p({\mathbf s})\,d{\mathbf s}\,, \label{96-nom} 
\end{equation}
which follows from the definitions (\ref{84-nom}) and (\ref{90-nom}). Next, the ``product of limits equal to limits of product'' rule gives the equality
\begin{equation}
\left[\lim_{{\mathbf s}_1\,\rightarrow\,\mu_1{\mathbf a}}\left\{+\,{\mathbf q}(\eta_{{\mathbf a}{\mathbf s}_1},\,{\mathbf r}_1)\right\}\right]\left[\lim_{{\mathbf s}_2\,\rightarrow\,\mu_2{\mathbf b}}\left\{-\,{\mathbf q}(\eta_{{\mathbf s}_2{\mathbf b}},\,{\mathbf r}_2)\right\}\right]=\lim_{\substack{{\mathbf s}_1\,\rightarrow\,\mu_1{\mathbf a} \\ {\mathbf s}_2\,\rightarrow\,\mu_2{\mathbf b}}}\Big\{-{\mathbf q}(\eta_{{\mathbf a}{\mathbf s}_1},\,{\mathbf r}_1)\,{\mathbf q}(\eta_{{\mathbf s}_2{\mathbf b}},\,{\mathbf r}_2)\Big\}. \label{limitproduct}
\end{equation}
This equality holds for quaternions or bivectors. It is proved explicitly in Appendix~\ref{Appen5} below. Moreover, in the Appendix~\ref{Appen4} the following equality is proved, which follows from the fact that $S^3$ as a set of unit quaternions remains closed under multiplication: 
\begin{equation}
{\mathbf q}(\eta_{{\mathbf a}{\mathbf s}_1},\,{\mathbf r}_{1}){\mathbf q}(\eta_{{\mathbf s}_2{\mathbf b}},\,{\mathbf r}_{2})={\mathbf q}(\eta_{{\mathbf u}{\mathbf v}},\,{\mathbf r}_{0})=\cos(\eta_{{\mathbf u}{\mathbf v}}) + (I_3{\mathbf r}_0)\,\sin(\eta_{{\mathbf u}{\mathbf v}})\,, \label{q-eq-appen}
\end{equation}
where
\begin{equation}
\eta_{{\mathbf u}{\mathbf v}}({\mathbf s}_1,\,{\mathbf s}_2)=\cos^{-1}\!\big\{({\mathbf a}\cdot{\mathbf s}_1)({\mathbf s}_2\cdot{\mathbf b})-({\mathbf a}\cdot{\mathbf s}_2)({\mathbf s}_1\cdot{\mathbf b})+({\mathbf a}\cdot{\mathbf b})({\mathbf s}_1\cdot{\mathbf s}_2)\big\} \label{38-22}
\end{equation}
and
\begin{equation}
{\mathbf r}_{0}({\mathbf s}_1,\,{\mathbf s}_2)=\frac{({\mathbf a}\cdot{\mathbf s}_1)({\mathbf s}_2\times{\mathbf b})+({\mathbf s}_2\cdot{\mathbf b})({\mathbf a}\times{\mathbf s}_1)-({\mathbf a}\times{\mathbf s}_1)\times({\mathbf s}_2\times{\mathbf b})}{\sin\left(\eta_{{\mathbf a}{\mathbf b}}\right)}. \label{101exam}
\end{equation}
Now, it is easy to see from (\ref{101exam}) that during the simultaneous detection processes by Alice and Bob, which are mathematically represented by the limits defined in (\ref{84-nom}) and (\ref{90-nom}), the axis vector ${\mathbf r}_{0}({\mathbf s}_1,\,{\mathbf s}_2)$ of the quaternion (\ref{q-eq-appen}) reduces to the zero vector $\vec{\mathbf 0}$,
\begin{equation}
\lim_{\substack{{\mathbf s}_1\,\rightarrow\,\mu_1{\mathbf a} \\ {\mathbf s}_2\,\rightarrow\,\mu_2{\mathbf b}}}{\mathbf r}_{0}({\mathbf s}_1,\,{\mathbf s}_2)=\vec{\mathbf 0}, \label{21-n}
\end{equation}
regardless of whether ${\mathbf s}_1 \!={\mathbf s}_2$ necessitated by the conservation of angular momentum holds. Thus the only part of the quaternion (\ref{q-eq-appen}) that survives in the simultaneous detection processes is its scalar part, which reduces to one of the following two possibilities:
\begin{numcases}{\lim_{\substack{{\mathbf s}_1\,\rightarrow\,\mu_1{\mathbf a} \\ {\mathbf s}_2\,\rightarrow\,\mu_2{\mathbf b}}}\,\left[\cos{\{\eta_{{\mathbf u}{\mathbf v}}({\mathbf s}_1,\,{\mathbf s}_2)\}}\right]=}
\mu_1\mu_2 = \text{sign}(\mathbf{a}\cdot\mathbf{s}^i_1)\,\text{sign}(\mathbf{b}\cdot\mathbf{s}^i_2) &  $ \text{if}\;\,{\mathbf s}^f_1\not=\,{\mathbf s}^f_2 $ \label{101--nnmm} \\ \notag \\
\cos(\eta_{{\mathbf a}{\mathbf b}}) & $ \text{if}\;\,{\mathbf s}^f_1=\,{\mathbf s}^f_2, $ \label{102--nnmm}
\end{numcases}
where the superscript $f$ on ${\mathbf s}$ indicates its direction just before detection. It is easy to see that these possibilities follow from (\ref{38-22}):
\begin{equation}
\lim_{\substack{{\mathbf s}_1\,\rightarrow\,\mu_1{\mathbf a} \\ {\mathbf s}_2\,\rightarrow\,\mu_2{\mathbf b}}}\,\left[\cos{\{\eta_{{\mathbf u}{\mathbf v}}({\mathbf s}_1,\,{\mathbf s}_2)\}}\right]=\lim_{\substack{{\mathbf s}_1\,\rightarrow\,\mu_1{\mathbf a} \\ {\mathbf s}_2\,\rightarrow\,\mu_2{\mathbf b}}}
\big\{({\mathbf a}\cdot{\mathbf s}_1)({\mathbf s}_2\cdot{\mathbf b})-({\mathbf a}\cdot{\mathbf s}_2)({\mathbf s}_1\cdot{\mathbf b})+({\mathbf a}\cdot{\mathbf b})({\mathbf s}_1\cdot{\mathbf s}_2)\big\}.
\end{equation}
Now, from (\ref{38-22}) and (\ref{102--nnmm}) we see that for ${\mathbf s}_1 = {\mathbf s}_2$ the angle $\eta_{{\mathbf u}{\mathbf v}}$ reduces to the angle $\eta_{{\mathbf a}{\mathbf b}}$ between ${\mathbf a}$ and ${\mathbf b}$, and thus (\ref{q-eq-appen}) reduces to 
\begin{equation}
{\mathbf q}(\eta_{{\mathbf a}{\mathbf b}},\,{\mathbf r}_{0})=\cos(\eta_{{\mathbf a}{\mathbf b}}) + (I_3{\mathbf r}_0)\,\sin(\eta_{{\mathbf a}{\mathbf b}}). \label{102forappen}
\end{equation}
As we will soon see, this is what reproduces the strong singlet correlations within $S^3$. But first it is instructive to understand the meaning and significance of (\ref{101--nnmm}), which describes the case in which the condition ${\mathbf s}^f_1=\,{\mathbf s}^f_2$ required by the conservation of spin angular momentum does not hold just before the detection processes begin. That is possible if some physical interaction external to the singlet system influences the rotation directions ${\mathbf s}_1$ and ${\mathbf s}_2$ sufficiently during their journey from the source to detectors to deviate them from each other. This destroys the fragile singlet system and therefore the correlations (\ref{96-nom}) fail to be sinusoidal:
\begin{align}
{\cal E}_{\mathrm{L.R.}}({\mathbf a},{\mathbf b})&=\int_{S^2}{\mathscr A}({\mathbf a},{\mathbf s}_1)\;{\mathscr B}({\mathbf b},{\mathbf s}_2)\,p({\mathbf s})\,d{\mathbf s} \label{22-nnmm}\\
&=\int_{S^2}\left[\lim_{{\mathbf s}_1\,\rightarrow\,\mu_1{\mathbf a}}\left\{+\,{\mathbf q}(\eta_{{\mathbf a}{\mathbf s}_1},\,{\mathbf r}_1)\right\}\right]\left[\lim_{{\mathbf s}_2\,\rightarrow\,\mu_2{\mathbf b}}\left\{-\,{\mathbf q}(\eta_{{\mathbf s}_2{\mathbf b}},\,{\mathbf r}_2)\right\}\right]p({\mathbf s})\,d{\mathbf s} \\
&=\int_{S^2}\Bigg[\lim_{\substack{{\mathbf s}_1\,\rightarrow\,\mu_1{\mathbf a} \\ {\mathbf s}_2\,\rightarrow\,\mu_2{\mathbf b}}}\Big\{-{\mathbf q}(\eta_{{\mathbf a}{\mathbf s}_1},\,{\mathbf r}_1)\,{\mathbf q}(\eta_{{\mathbf s}_2{\mathbf b}},\,{\mathbf r}_2)\Big\}\Bigg]p({\mathbf s})\,d{\mathbf s} \label{5forappenmm} \\
&=\int_{S^2}\Bigg[\lim_{\substack{{\mathbf s}_1\,\rightarrow\,\mu_1{\mathbf a} \\ {\mathbf s}_2\,\rightarrow\,\mu_2{\mathbf b}}}\Big\{-{\mathbf q}(\eta_{{\mathbf u}{\mathbf v}},\,{\mathbf r}_{0})\Big\} \Bigg]p({\mathbf s})\,d{\mathbf s} \label{6forappenmm} \\
&=\int_{S^2}\Bigg[\lim_{\substack{{\mathbf s}_1\,\rightarrow\,\mu_1{\mathbf a} \\ {\mathbf s}_2\,\rightarrow\,\mu_2{\mathbf b}}}\Big\{-\cos(\,\eta_{{\mathbf u}{\mathbf v}})-\left(I_3{\mathbf r}_0\right)\sin(\,\eta_{{\mathbf u}{\mathbf v}})\Big\}\!\Bigg]p({\mathbf s})\,d{\mathbf s} \\
&=\int_{S^2}\Bigg[\lim_{\substack{{\mathbf s}_1\,\rightarrow\,\mu_1{\mathbf a} \\ {\mathbf s}_2\,\rightarrow\,\mu_2{\mathbf b}}}\Big\{-\cos(\,\eta_{{\mathbf u}{\mathbf v}})\Big\}-\lim_{\substack{{\mathbf s}_1\,\rightarrow\,\mu_1{\mathbf a} \\ {\mathbf s}_2\,\rightarrow\,\mu_2{\mathbf b}}}\Big\{I_3{\mathbf r}_0\Big\}\times\lim_{\substack{{\mathbf s}_1\,\rightarrow\,\mu_1{\mathbf a} \\ {\mathbf s}_2\,\rightarrow\,\mu_2{\mathbf b}}}\Big\{\sin(\,\eta_{{\mathbf u}{\mathbf v}})\Big\}\Bigg]\,p({\mathbf s})\,d{\mathbf s} \\
&=\int_{S^2}\Bigg[\big\{-\mu_1\mu_2\big\}-\left\{I_3\vec{\mathbf 0}\right\}\times\lim_{\substack{{\mathbf s}_1\,\rightarrow\,\mu_1{\mathbf a} \\ {\mathbf s}_2\,\rightarrow\,\mu_2{\mathbf b}}}\Big\{\sin(\,\eta_{{\mathbf u}{\mathbf v}})\Big\}\Bigg]p({\mathbf s})\;d{\mathbf s} \\
&=\int_{S^2}\Bigg[\Big\{\!-\text{sign}(\mathbf{a}\cdot\mathbf{s}^i_1)\,\text{sign}(\mathbf{b}\cdot\mathbf{s}^i_2)\Big\}-0\Bigg]p({\mathbf s})\;d{\mathbf s} \;\;[\text{the additive identity 0 is the same across all grades in}\;\mathrm{Cl_{3,0}}],
\end{align}
where the last equation is well known to produce correlations that are incapable of exceeding the bounds on Bell inequalities \cite{Bell-1964,Peres}:
\begin{equation}
{\cal E}_{\mathrm{L.R.}}({\mathbf a},{\mathbf b})\,=\int_{S^2}\!\left\{-\,\text{sign}(\mathbf{a}\cdot\mathbf{s}^i_1)\;\text{sign}(\mathbf{b}\cdot\mathbf{s}^i_2)\right\}\,p({\mathbf s})\;d{\mathbf s} \,=\,
\begin{cases}
-\,1\,+\,\frac{2}{\pi}\,\eta_{{\mathbf a}{\mathbf b}}
\;\;\;\text{if} &\!\! 0 \leqslant \eta_{{\mathbf a}{\mathbf b}} \leqslant \pi, \\
\\
+\,3\,-\,\frac{2}{\pi}\,\eta_{{\mathbf a}{\mathbf b}}
\;\;\;\text{if} &\!\! \pi \leqslant \eta_{{\mathbf a}{\mathbf b}} \leqslant 2\pi.
\end{cases} \label{twoobserve}
\end{equation}
\begin{figure}
\scalebox{1}{
\begin{pspicture}(4.5,-0.7)(-4.5,5.7)
\psset{xunit=0.5mm,yunit=4cm}
\psaxes[axesstyle=frame,linewidth=0.01mm,tickstyle=full,ticksize=0pt,dx=90\psxunit,Dx=180,dy=1
\psyunit,Dy=+2,Oy=-1](0,0)(180,1.0)
\psline[linewidth=0.2mm,arrowinset=0.3,arrowsize=2pt 3,arrowlength=2]{->}(0,0.5)(190,0.5)
\psline[linewidth=0.2mm]{-}(45,0)(45,1)
\psline[linewidth=0.2mm]{-}(90,0)(90,1)
\psline[linewidth=0.2mm]{-}(135,0)(135,1)
\psline[linewidth=0.2mm,arrowinset=0.3,arrowsize=2pt 3,arrowlength=2]{->}(0,0)(0,1.2)
\psline[linewidth=0.35mm,linestyle=dashed]{-}(0,0)(90,1)
\psline[linewidth=0.35mm,linestyle=dashed]{-}(90,1)(180,0)
\put(2.1,-0.38){${90}$}
\put(6.5,-0.38){${270}$}
\put(-0.63,3.92){${+}$}
\put(-0.7,5.0){{\large ${{\cal E}_{{\!}_{\mathrm{L.R.}}}}$}${\!({\bf a},{\bf b})}$}
\put(-0.38,1.93){${0}$}
\put(9.65,1.85){\large ${\eta_{{\bf a}{\bf b}}}$}
\psplot[linewidth=0.35mm,linecolor=black]{0.0}{180}{x dup cos exch cos mul 1.0 mul neg 1 add}
\end{pspicture}}
\caption{Graphs of the correlations (\ref{twoobserve}) and (\ref{whichr}). The x-axis depicts the angle in degrees between the detector directions ${\mathbf a}$ and ${\mathbf b}$, and the y-axis depicts the corresponding expectation value or correlation. The dotted straight lines depict the correlations if ${\mathbf s}^f_1\not=\,{\mathbf s}^f_2$ holds, and the solid curve represents the predictions of the 3-sphere model for which ${\mathbf s}^f_1=\,{\mathbf s}^f_2$ holds. After~Christian \cite{IEEE-2}.}
\label{fig-6}
\end{figure}
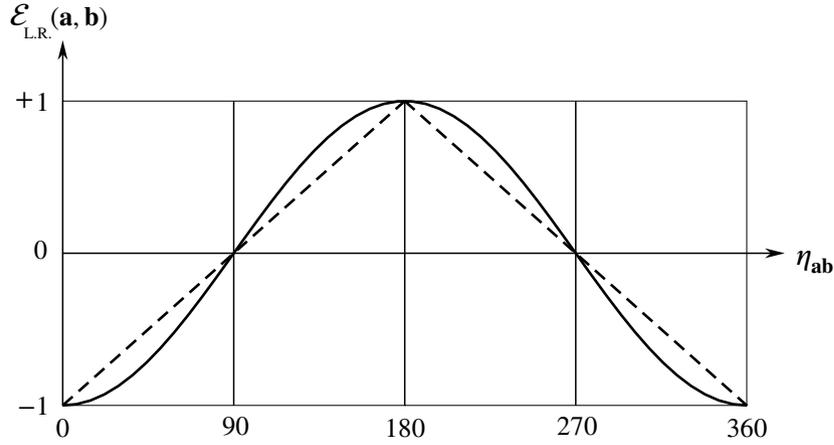
In Fig.~\ref{fig-6}\!\!\! these weak correlations are depicted by saw-tooth shaped dashed straight lines. By contrast, the sinusoidal solid curve in Fig.~\ref{fig-6}\!\!\! depicts the strong singlet correlations predicted by the 3-sphere model and quantum mechanics. They follow for the result (\ref{102--nnmm}) in which the spin angular momentum is conserved up until the detection processes, and can be derived as follows:
\begin{align}
{\cal E}_{\mathrm{L.R.}}({\mathbf a},{\mathbf b})&=\int_{S^2}{\mathscr A}({\mathbf a},{\mathbf s}_1)\;{\mathscr B}({\mathbf b},{\mathbf s}_2)\,p({\mathbf s})\,d{\mathbf s} \label{22-nn}\\
&=\int_{S^2}\left[\lim_{{\mathbf s}_1\,\rightarrow\,\mu_1{\mathbf a}}\left\{+\,{\mathbf q}(\eta_{{\mathbf a}{\mathbf s}_1},\,{\mathbf r}_1)\right\}\right]\left[\lim_{{\mathbf s}_2\,\rightarrow\,\mu_2{\mathbf b}}\left\{-\,{\mathbf q}(\eta_{{\mathbf s}_2{\mathbf b}},\,{\mathbf r}_2)\right\}\right]p({\mathbf s})\,d{\mathbf s} \\
&=\int_{S^2}\Bigg[\lim_{\substack{{\mathbf s}_1\,\rightarrow\,\mu_1{\mathbf a} \\ {\mathbf s}_2\,\rightarrow\,\mu_2{\mathbf b}}}\Big\{-{\mathbf q}(\eta_{{\mathbf a}{\mathbf s}_1},\,{\mathbf r}_1)\,{\mathbf q}(\eta_{{\mathbf s}_2{\mathbf b}},\,{\mathbf r}_2)\Big\}\Bigg]p({\mathbf s})\,d{\mathbf s} \label{5forappen} \\
&=\int_{S^2}\Bigg[\lim_{\substack{{\mathbf s}_1\,\rightarrow\,\mu_1{\mathbf a} \\ {\mathbf s}_2\,\rightarrow\,\mu_2{\mathbf b}}}\Big\{-{\mathbf q}(\eta_{{\mathbf a}{\mathbf b}},\,{\mathbf r}_{0})\Big\} \Bigg]p({\mathbf s})\,d{\mathbf s} \label{6forappen} \\
&=\int_{S^2}\Bigg[\lim_{\substack{{\mathbf s}_1\,\rightarrow\,\mu_1{\mathbf a} \\ {\mathbf s}_2\,\rightarrow\,\mu_2{\mathbf b}}}\Big\{-\cos(\,\eta_{{\mathbf a}{\mathbf b}})-\left(I_3{\mathbf r}_0\right)\sin(\,\eta_{{\mathbf a}{\mathbf b}})\Big\}\!\Bigg]p({\mathbf s})\,d{\mathbf s} \\
&=\,-\cos(\,\eta_{{\mathbf a}{\mathbf b}})-\int_{S^2}\Bigg[\lim_{\substack{{\mathbf s}_1\,\rightarrow\,\mu_1{\mathbf a} \\ {\mathbf s}_2\,\rightarrow\,\mu_2{\mathbf b}}}\left(I_3{\mathbf r}_0\right)\Bigg]\sin(\,\eta_{{\mathbf a}{\mathbf b}})\,p({\mathbf s})\,d{\mathbf s} \\
&=\,-\cos(\,\eta_{{\mathbf a}{\mathbf b}})-\left[I_3\vec{\mathbf 0}\right]\sin(\,\eta_{{\mathbf a}{\mathbf b}})\int_{S^2}\!p({\mathbf s})\;d{\mathbf s} \\
&=\,-\cos(\,\eta_{{\mathbf a}{\mathbf b}})-\left[I_3\vec{\mathbf 0}\right]\sin(\,\eta_{{\mathbf a}{\mathbf b}}) \\
&=-\cos(\,\eta_{{\mathbf a}{\mathbf b}})-0\;\;\;[\text{the additive identity 0 is the same across all grades in}\;\mathrm{Cl_{3,0}}]\,, \label{whichr}
\end{align}
where the equality between the product of two quaternions appearing in (\ref{5forappen}) and the one in (\ref{6forappen}) is proved in Appendix~\ref{Appen4}. The only difference between the derivations of the weak correlations (\ref{twoobserve}) and strong correlations (\ref{whichr}) is the condition ${\mathbf s}^f_1=\,{\mathbf s}^f_2$.

It is important to bear in mind that the above local-realistic derivation of the singlet correlations is a {\it theoretical} prediction of the quaternionic 3-sphere model \cite{IEEE-1,IEEE-2}. It should not be confused with after-the-event data analyses followed by experimenters \cite{IEEE-4}.

Note also that there are only two differences between the above derivation of the singlet correlations and the one I discussed in Subsection~\ref{Sec-2.1} above. First, I have not used the relative handedness $\lambda$ between the spin bivectors and the detector bivectors. This eliminates one of the concerns raised in Lasenby \cite{Lasenby-AGACSE-2021}, even though I demonstrated in Subsection~\ref{Las-B} that that concern is unjustified. Second, instead of $\lambda$, I have introduced and used the limits discussed just below Eq.~(\ref{79-nom}). These limits have quite transparent meanings. As we see in (\ref{92-nom}), they unambiguously predict ${\mathscr A}{\mathscr B}=\pm1$ in general for the choices ${\mathbf a}\not={\mathbf b}$, even when ${\mathbf s}_1 \not= {\mathbf s}_2$ holds.

Finally, the derivation presented above is independently verified in Diether\cite{Diether} in a numerical simulation using Mathematica. In Christian\cite{IJTP,Symmetric} I have also proposed a macroscopic experiment that may be able to falsify the 3-sphere hypothesis. For this purpose, let us note again that when the zero spins angular momentum is conserved, as it must be, the angular relation (\ref{38-22}) reduces to
\begin{equation}
\cos(\eta_{{\mathbf u}{\mathbf v}})=({\mathbf a}\cdot{\mathbf s}_1)({\mathbf s}_2\cdot{\mathbf b})-({\mathbf a}\cdot{\mathbf s}_2)({\mathbf s}_1\cdot{\mathbf b})+({\mathbf a}\cdot{\mathbf b})({\mathbf s}_1\cdot{\mathbf s}_2)=\cos(\eta_{{\mathbf a}{\mathbf b}}).
\end{equation}
Now, the second equality in the above equation can be equivalently written as
\begin{equation}
-\int_{S^2}\left[\lim_{\substack{{\mathbf s}_1\,\rightarrow\,\mu_1{\mathbf a} \\ {\mathbf s}_2\,\rightarrow\,\mu_2{\mathbf b}}}\left\{({\mathbf a}\cdot{\mathbf s}_1)({\mathbf s}_2\cdot{\mathbf b})-({\mathbf a}\cdot{\mathbf s}_2)({\mathbf s}_1\cdot{\mathbf b})+({\mathbf a}\cdot{\mathbf b})({\mathbf s}_1\cdot{\mathbf s}_2)\right\}\right]p({\mathbf s})\;d{\mathbf s}\;=\,-\int_{S^2}\left[\lim_{\substack{{\mathbf s}_1\,\rightarrow\,\mu_1{\mathbf a} \\ {\mathbf s}_2\,\rightarrow\,\mu_2{\mathbf b}}}\left\{
\cos(\eta_{{\mathbf a}{\mathbf b}})\right\}\right]p({\mathbf s})\;d{\mathbf s}\,,
\end{equation}
which, upon taking limits, simplifies to
\begin{equation}
\int_{S^2}\!\left\{-\mu_1\mu_2\right\}\,p({\mathbf s})\;d{\mathbf s}=-\cos(\eta_{{\mathbf a}{\mathbf b}}).
\end{equation}
Using (\ref{85-nom}), (\ref{91-nom}), and definitions (\ref{79-nom}), we recognize that this equation is equivalent to the prediction (\ref{whichr}) of the 3-sphere model:
\begin{equation}
{\cal E}_{\mathrm{L.R.}}({\mathbf a},{\mathbf b})\,=\int_{S^2}\!\left\{-\,\text{sign}(\mathbf{a}\cdot\mathbf{s}^i_1)\;\text{sign}(\mathbf{b}\cdot\mathbf{s}^i_2)\right\}\,p({\mathbf s})\;d{\mathbf s}\approx\lim_{\,n\,\gg\,1}\!\left[\frac{1}{n}\!\sum_{k\,=\,1}^{n}
\left\{-\,\text{sign}({\bf a}\cdot{\bf s}^k_1)\;
\text{sign}({\bf b}\cdot{\bf s}^k_2)\right\}\right]
=-\cos(\eta_{{\mathbf a}{\mathbf b}}). \label{127-yes}
\end{equation}
It is this prediction (\ref{127-yes}) that can be tested experimentally in a macroscopic setting, as I have explained in detail in Christian \cite{Symmetric}. Note that the summation in (\ref{127-yes}) involves sign functions of ordinary vectors, which can be easily measured and calculated. In Christian \cite{Symmetric} I have also provided a simplified derivation of the correlation (\ref{127-yes}) that does not require using the orientation of $S^3$.

\section*{Conflict of Interest}
The author declares no conflict of interests in this work.

\section*{ORCID}
{\it Joy Christian} \url{https://orcid.org/0000-0002-8741-6943}

\appendix

\counterwithin{equation}{section}

\section{Proofs of Lemmas 2, 3, and 4, and of Equations (96) and (105)}

\subsection{Proof of Lemma~\ref{L2}} \label{Appen1}

To verify the multiplicativity (\ref{no-5}) of the norms $||{\mathbb Q}_{z}||$, consider the left-hand side of (\ref{no-5}) for two distinct elements of ${{\cal K}^{\lambda}}$:
\begin{equation}
{\mathbb Q}_{z1} =\, {\bf q}_{r1} + {\bf q}_{d1}\,\varepsilon\;\;\;\;\text{and}\;\;\;\;{\mathbb Q}_{z2} =\, {\bf q}_{r2} + {\bf q}_{d2}\,\varepsilon\,.
\end{equation}
Since the pseudoscalar $\varepsilon$ commutes with every element of ${\cal K}^{\lambda}$, the geometric product of the above elements gives  
\begin{equation}
{\mathbb Q}_{z1}\,{\mathbb Q}_{z2}\,=\;\left({\bf q}_{r1}\,{\bf q}_{r2}\,+\,{\bf q}_{d1}\,{\bf q}_{d2}\right)\,+\,\left({\bf q}_{r1}\,{\bf q}_{d2}\,+\,{\bf q}_{d1}\,{\bf q}_{r2}\right)\,\varepsilon\,.  \label{new24}
\end{equation}
where I have used ${\varepsilon^2=1}$. If we now use again the fact that ${\varepsilon}$, along with ${\varepsilon^{\dagger}=\varepsilon}$ and ${\varepsilon^2=1}$, commutes with every element of ${{\cal K}^{\lambda}}$ and consequently with all ${{\bf q}_r}$, ${{\bf q}^{\dagger}_r}$, ${{\bf q}_d}$ and ${{\bf q}^{\dagger}_d}$, and work out ${{\mathbb Q}^{\dagger}_{z1}}$, ${{\mathbb Q}^{\dagger}_{z2}}$ and the products ${{\mathbb Q}_{z1}{\mathbb Q}^{\dagger}_{z1}}$, ${\,{\mathbb Q}_{z2}{\mathbb Q}^{\dagger}_{z2}}$ and ${({\mathbb Q}_{z1}\,{\mathbb Q}_{z2})^{\dagger}}$ as
\begin{align}
{\mathbb Q}^{\dagger}_{z1} &=\, {\bf q}^{\dagger}_{r1} + {\bf q}^{\dagger}_{d1}\,\varepsilon\,, \\
{\mathbb Q}^{\dagger}_{z2} &=\, {\bf q}^{\dagger}_{r2} + {\bf q}^{\dagger}_{d2}\,\varepsilon\,, \\
{\mathbb Q}_{z1}\,{\mathbb Q}^{\dagger}_{z1} &=\,\left({\bf q}_{r1}\,{\bf q}^{\dagger}_{r1}\,+\,{\bf q}_{d1}\,{\bf q}^{\dagger}_{d1}\right)\,+\,\left({\bf q}_{r1}\,{\bf q}^{\dagger}_{d1}\,+\,{\bf q}_{d1}\,{\bf q}^{\dagger}_{r1}\right)\,\varepsilon\,, \label{t-39} \\
{\mathbb Q}_{z2}\,{\mathbb Q}^{\dagger}_{z2} &=\,\left({\bf q}_{r2}\,{\bf q}^{\dagger}_{r2}\,+\,{\bf q}_{d2}\,{\bf q}^{\dagger}_{d2}\right)\,+\,\left({\bf q}_{r2}\,{\bf q}^{\dagger}_{d2}\,+\,{\bf q}_{d2}\,{\bf q}^{\dagger}_{r2}\right)\,\varepsilon\,, \label{t-40} \\
\text{and}\;\;\;({\mathbb Q}_{z1}\,{\mathbb Q}_{z2})^{\dagger}\,&=\;\left({\bf q}^{\dagger}_{r2}\,{\bf q}^{\dagger}_{r1}\,+\,{\bf q}^{\dagger}_{d2}\,{\bf q}^{\dagger}_{d1}\right)\,+\,\left({\bf q}^{\dagger}_{d2}\,{\bf q}^{\dagger}_{r1}\,+\,{\bf q}^{\dagger}_{r2}\,{\bf q}^{\dagger}_{d1}\right)\,\varepsilon\,, \label{new30}
\end{align}
then, using (\ref{new24}) to (\ref{new30}), the norm relation (\ref{no-5}) is not difficult to verify. To that end, let us work out the geometric product ${({\mathbb Q}_{z1}\,{\mathbb Q}_{z2})({\mathbb Q}_{z1}\,{\mathbb Q}_{z2})^{\dagger}}$ using the expressions (\ref{new24}) and (\ref{new30}), which gives
\begin{align}
({\mathbb Q}_{z1}\,{\mathbb Q}_{z2})({\mathbb Q}_{z1}\,{\mathbb Q}_{z2})^{\dagger} &=\left\{\left({\bf q}_{r1}\,{\bf q}_{r2}+{\bf q}_{d1}\,{\bf q}_{d2}\right)
\left({\bf q}^{\dagger}_{r2}\,{\bf q}^{\dagger}_{r1}+{\bf q}^{\dagger}_{d2}\,{\bf q}^{\dagger}_{d1}\right)
+\left({\bf q}_{r1}\,{\bf q}_{d2}+{\bf q}_{d1}\,{\bf q}_{r2}\right)
\left({\bf q}^{\dagger}_{d2}\,{\bf q}^{\dagger}_{r1}+{\bf q}^{\dagger}_{r2}\,{\bf q}^{\dagger}_{d1}\right)\right\} \notag \\
&\;\;\;\;\;\;+\left\{\left({\bf q}_{r1}\,{\bf q}_{d2}+{\bf q}_{d1}\,{\bf q}_{r2}\right)
\left({\bf q}^{\dagger}_{r2}\,{\bf q}^{\dagger}_{r1}+{\bf q}^{\dagger}_{d2}\,{\bf q}^{\dagger}_{d1}\right)
+\left({\bf q}_{r1}\,{\bf q}_{r2}+{\bf q}_{d1}\,{\bf q}_{d2}\right)
\left({\bf q}^{\dagger}_{d2}\,{\bf q}^{\dagger}_{r1}+{\bf q}^{\dagger}_{r2}\,{\bf q}^{\dagger}_{d1}\right)\right\}\varepsilon\,. \label{prod}
\end{align}
Using (\ref{new24}), (\ref{new30}), and their product evaluated in (\ref{prod}), we can now evaluate the left-hand side of (\ref{5}) as follows:
\begin{align}
||{\mathbb Q}_{z1}\,{\mathbb Q}_{z2}||^2 &= \left({\mathbb Q}_{z1}\,{\mathbb Q}_{z2}\right)\left({\mathbb Q}_{z1}\,{\mathbb Q}_{z2}\right)^{\dagger} \\
&=\Big\{\left({\bf q}_{r1}\,{\bf q}_{r2}+{\bf q}_{d1}\,{\bf q}_{d2}\right) +\left({\bf q}_{r1}\,{\bf q}_{d2}+{\bf q}_{d1}\,{\bf q}_{r2}\right)\varepsilon\Big\}\left\{\left({\bf q}^{\dagger}_{r2}\,{\bf q}^{\dagger}_{r1}+{\bf q}^{\dagger}_{d2}\,{\bf q}^{\dagger}_{d1}\right)+\left({\bf q}^{\dagger}_{d2}\,{\bf q}^{\dagger}_{r1}+{\bf q}^{\dagger}_{r2}\,{\bf q}^{\dagger}_{d1}\right)\varepsilon\right\} \\ 
&=\left\{\left({\bf q}_{r1}\,{\bf q}_{r2}+{\bf q}_{d1}\,{\bf q}_{d2}\right)
\left({\bf q}^{\dagger}_{r2}\,{\bf q}^{\dagger}_{r1}+{\bf q}^{\dagger}_{d2}\,{\bf q}^{\dagger}_{d1}\right)
+\left({\bf q}_{r1}\,{\bf q}_{d2}+{\bf q}_{d1}\,{\bf q}_{r2}\right)
\left({\bf q}^{\dagger}_{d2}\,{\bf q}^{\dagger}_{r1}+{\bf q}^{\dagger}_{r2}\,{\bf q}^{\dagger}_{d1}\right)\right\} \notag \\
&\;\;\;\;\;\;\;\,+\left\{\left({\bf q}_{r1}\,{\bf q}_{r2}+{\bf q}_{d1}\,{\bf q}_{d2}\right)
\left({\bf q}^{\dagger}_{d2}\,{\bf q}^{\dagger}_{r1}+{\bf q}^{\dagger}_{r2}\,{\bf q}^{\dagger}_{d1}\right)+\left({\bf q}_{r1}\,{\bf q}_{d2}+{\bf q}_{d1}\,{\bf q}_{r2}\right)
\left({\bf q}^{\dagger}_{r2}\,{\bf q}^{\dagger}_{r1}+{\bf q}^{\dagger}_{d2}\,{\bf q}^{\dagger}_{d1}\right)
\right\}\varepsilon \\
&=\Big\{{\bf q}_{r1}\,{\bf q}_{r2}\,{\bf q}^{\dagger}_{r2}\,{\bf q}^{\dagger}_{r1}\,+\, {\bf q}_{r1}\,{\bf q}_{r2}\,{\bf q}^{\dagger}_{d2}\,{\bf q}^{\dagger}_{d1}
\,+\, {\bf q}_{d1}\,{\bf q}_{d2}\,{\bf q}^{\dagger}_{r2}\,{\bf q}^{\dagger}_{r1} \,+\, {\bf q}_{d1}\,{\bf q}_{d2}\,{\bf q}^{\dagger}_{d2}\,{\bf q}^{\dagger}_{d1} \notag \\
&\,\;\;\;\;\;\;\;+\, {\bf q}_{r1}\,{\bf q}_{d2}\,{\bf q}^{\dagger}_{d2}\,{\bf q}^{\dagger}_{r1} \,+\, 
{\bf q}_{r1}\,{\bf q}_{d2}\,{\bf q}^{\dagger}_{r2}\,{\bf q}^{\dagger}_{d1} \,+\,{\bf q}_{d1}\,{\bf q}_{r2}\,{\bf q}^{\dagger}_{d2}\,{\bf q}^{\dagger}_{r1} \,+\, {\bf q}_{d1}\,{\bf q}_{r2}\,{\bf q}^{\dagger}_{r2}\,{\bf q}^{\dagger}_{d1}\Big\} \notag \\
&\;\;\;\;\;\;\;\;\;\;\;+\Big\{{\bf q}_{r1}\,{\bf q}_{r2}\,{\bf q}^{\dagger}_{d2}\,{\bf q}^{\dagger}_{r1} \,+\, {\bf q}_{r1}\,{\bf q}_{r2}\,{\bf q}^{\dagger}_{r2}\,{\bf q}^{\dagger}_{d1} \,+\, {\bf q}_{d1}\,{\bf q}_{d2}\,{\bf q}^{\dagger}_{d2}\,{\bf q}^{\dagger}_{r1} \,+\, {\bf q}_{d1}\,{\bf q}_{d2}\,{\bf q}^{\dagger}_{r2}\,{\bf q}^{\dagger}_{d1} \notag \\
&\,\;\;\;\;\;\;\;\;\;\;\;\;\;\;\;+\, {\bf q}_{r1}\,{\bf q}_{d2}\,{\bf q}^{\dagger}_{r2}\,{\bf q}^{\dagger}_{r1} \,+\, {\bf q}_{r1}\,{\bf q}_{d2}\,{\bf q}^{\dagger}_{d2}\,{\bf q}^{\dagger}_{d1} \,+\,
{\bf q}_{d1}\,{\bf q}_{r2}\,{\bf q}^{\dagger}_{r2}\,{\bf q}^{\dagger}_{r1} \,+\, {\bf q}_{d1}\,{\bf q}_{r2}\,{\bf q}^{\dagger}_{d2}\,{\bf q}^{\dagger}_{d1}\Big\}\,\varepsilon \\
&=\Big\{\varrho^2_{r1}\,\varrho^2_{r2}\,+\, {\bf q}_{r1}\,{\bf q}_{r2}\,{\bf q}^{\dagger}_{d2}\,{\bf q}^{\dagger}_{d1}
\,+\, {\bf q}_{d1}\,{\bf q}_{d2}\,{\bf q}^{\dagger}_{r2}\,{\bf q}^{\dagger}_{r1} \,+\,\varrho^2_{d1}\,\varrho^2_{d2} \notag \\
&\,\;\;\;\;\;\;\;+\, \varrho^2_{r1}\,\varrho^2_{d2}\,+\, 
{\bf q}_{r1}\,{\bf q}_{d2}\,{\bf q}^{\dagger}_{r2}\,{\bf q}^{\dagger}_{d1} \,+\,{\bf q}_{d1}\,{\bf q}_{r2}\,{\bf q}^{\dagger}_{d2}\,{\bf q}^{\dagger}_{r1} \,+\,\varrho^2_{d1}\,\varrho^2_{r2}\Big\} \notag \\
&\;\;\;\;\;\;\;\;\;\;\;+\Big\{{\bf q}_{r1}\,{\bf q}_{r2}\,{\bf q}^{\dagger}_{d2}\,{\bf q}^{\dagger}_{r1} \,+\, \varrho^2_{r2}\,{\bf q}_{r1}\,{\bf q}^{\dagger}_{d1} \,+\, \varrho^2_{d2}\,{\bf q}_{d1}\,{\bf q}^{\dagger}_{r1} \,+\, {\bf q}_{d1}\,{\bf q}_{d2}\,{\bf q}^{\dagger}_{r2}\,{\bf q}^{\dagger}_{d1} \notag \\
&\,\;\;\;\;\;\;\;\;\;\;\;\;\;\;\;+\, {\bf q}_{r1}\,{\bf q}_{d2}\,{\bf q}^{\dagger}_{r2}\,{\bf q}^{\dagger}_{r1} \,+\, \varrho^2_{d2}\,{\bf q}_{r1}\,{\bf q}^{\dagger}_{d1} \,+\,\varrho^2_{r2}\,{\bf q}_{d1}\,{\bf q}^{\dagger}_{r1} \,+\, {\bf q}_{d1}\,{\bf q}_{r2}\,{\bf q}^{\dagger}_{d2}\,{\bf q}^{\dagger}_{d1}\Big\}\,\varepsilon \\
&=\Big\{\!\left(\varrho^2_{r1}+\varrho^2_{d1}\right)\!\left(\varrho^2_{r2}+\varrho^2_{d2}\right)+{\bf q}_{r1}\left({\bf q}_{r2}\,{\bf q}^{\dagger}_{d2}+{\bf q}_{d2}\,{\bf q}^{\dagger}_{r2}\right){\bf q}^{\dagger}_{d1}+{\bf q}_{d1}\left({\bf q}_{r2}\,{\bf q}^{\dagger}_{d2}+
{\bf q}_{d2}\,{\bf q}^{\dagger}_{r2}\right){\bf q}^{\dagger}_{r1}\Big\} \notag \\
&\;\;\;\;\;\;\;\;\;\;\;+\Big\{{\bf q}_{r1}\left({\bf q}_{r2}\,{\bf q}^{\dagger}_{d2}+{\bf q}_{d2}\,{\bf q}^{\dagger}_{r2}\right){\bf q}^{\dagger}_{r1} \,+\, \varrho^2_{r2}\left({\bf q}_{r1}\,{\bf q}^{\dagger}_{d1}+{\bf q}_{d1}\,{\bf q}^{\dagger}_{r1}\right) \notag \\
&\;\;\;\;\;\;\;\;\;\;\;\;\;\;\;\;\;\;+\varrho^2_{d2}\left({\bf q}_{r1}\,{\bf q}^{\dagger}_{d1}+{\bf q}_{d1}\,{\bf q}^{\dagger}_{r1}\right) + {\bf q}_{d1}\left({\bf q}_{r2}\,{\bf q}^{\dagger}_{d2}+{\bf q}_{d2}\,{\bf q}^{\dagger}_{r2}\right){\bf q}^{\dagger}_{d1}\Big\}\,\varepsilon \\
&=\left\{\varrho^2_{r1}\,\varrho^2_{r2}+\varrho^2_{r1}\,\varrho^2_{d2}+\varrho^2_{d1}\,\varrho^2_{r2}+\varrho^2_{d1}\,\varrho^2_{d2} \,+\,\left({\bf q}_{r1}\,{\bf q}^{\dagger}_{d1}+{\bf q}_{d1}\,{\bf q}^{\dagger}_{r1}\right)\left({\bf q}_{r2}\,{\bf q}^{\dagger}_{d2}+{\bf q}_{d2}\,{\bf q}^{\dagger}_{r2}\right)\right\} \notag \\
&\;\;\;\;\;\;\;\;\;\;+\left\{\left( \varrho^2_{r1}\,+\,\varrho^2_{d1}\right) \left({\bf q}_{r2}\,{\bf q}^{\dagger}_{d2}+{\bf q}_{d2}\,{\bf q}^{\dagger}_{r2}\right)
+\left( \varrho^2_{r2}\,+\,\varrho^2_{d2}\right) \left({\bf q}_{r1}\,{\bf q}^{\dagger}_{d1}+{\bf q}_{d1}\,{\bf q}^{\dagger}_{r1}\right)\right\}\,\varepsilon\,. \label{B8}
\end{align}
Again, since the quantities such as ${\bf q}_{r1}\,{\bf q}^{\dagger}_{d1}+{\bf q}_{d1}\,{\bf q}^{\dagger}_{r1}$ are scalar quantities, we see that the above product also resembles a split complex number similar to ${\mathbb Q}_{z}\,{\mathbb Q}^{\dagger}_{z}$ in (\ref{10c}). In other words, it is a sum of a scalar and a pseudoscalar, as in (\ref{23a}).

Next, we can evaluate the right-hand side of the relation (\ref{5}) using (\ref{t-39}) and (\ref{t-40}) as follows:
\begin{align}
||{\mathbb Q}_{z1}||^2\,||{\mathbb Q}_{z2}||^2
&= \left({\mathbb Q}_{z1}\,{\mathbb Q}^{\dagger}_{z1}\right)\left({\mathbb Q}_{z2}\,{\mathbb Q}^{\dagger}_{z2}\right) \\
&=\left\{\left({\bf q}_{r1}\,{\bf q}^{\dagger}_{r1}+{\bf q}_{d1}\,{\bf q}^{\dagger}_{d1}\right)+\left({\bf q}_{r1}\,{\bf q}^{\dagger}_{d1}+{\bf q}_{d1}\,{\bf q}^{\dagger}_{r1}\right)\varepsilon\right\}\left\{\left({\bf q}_{r2}\,{\bf q}^{\dagger}_{r2}+{\bf q}_{d2}\,{\bf q}^{\dagger}_{d2}\right)+\left({\bf q}_{r2}\,{\bf q}^{\dagger}_{d2}+{\bf q}_{d2}\,{\bf q}^{\dagger}_{r2}\right)\varepsilon\right\} \\
&=\left\{\left({\bf q}_{r1}\,{\bf q}^{\dagger}_{r1}+{\bf q}_{d1}\,{\bf q}^{\dagger}_{d1}\right)\left({\bf q}_{r2}\,{\bf q}^{\dagger}_{r2}+{\bf q}_{d2}\,{\bf q}^{\dagger}_{d2}\right)+ \left({\bf q}_{r1}\,{\bf q}^{\dagger}_{d1}+{\bf q}_{d1}\,{\bf q}^{\dagger}_{r1}\right)\left({\bf q}_{r2}\,{\bf q}^{\dagger}_{d2}+{\bf q}_{d2}\,{\bf q}^{\dagger}_{r2}\right)
\right\} \notag \\
&\;\;\;\;\;\;\;\;\;\;+\Big\{\left({\bf q}_{r1}\,{\bf q}^{\dagger}_{r1}+{\bf q}_{d1}\,{\bf q}^{\dagger}_{d1}\right) \left({\bf q}_{r2}\,{\bf q}^{\dagger}_{d2}+{\bf q}_{d2}\,{\bf q}^{\dagger}_{r2}\right)+\left({\bf q}_{r1}\,{\bf q}^{\dagger}_{d1}+{\bf q}_{d1}\,{\bf q}^{\dagger}_{r1}\right)\left({\bf q}_{r2}\,{\bf q}^{\dagger}_{r2}+{\bf q}_{d2}\,{\bf q}^{\dagger}_{d2}\right)
\Big\}\,\varepsilon    \\
&=\left\{\left(\varrho^2_{r1}+\varrho^2_{d1}\right)\left(\varrho^2_{r2}+\varrho^2_{d2}\right) \,+\,\left({\bf q}_{r1}\,{\bf q}^{\dagger}_{d1}+{\bf q}_{d1}\,{\bf q}^{\dagger}_{r1}\right)\left({\bf q}_{r2}\,{\bf q}^{\dagger}_{d2}+{\bf q}_{d2}\,{\bf q}^{\dagger}_{r2}\right)\right\} \notag \\
&\;\;\;\;\;\;\;\;\;\;+\left\{\left( \varrho^2_{r1}\,+\,\varrho^2_{d1}\right) \left({\bf q}_{r2}\,{\bf q}^{\dagger}_{d2}+{\bf q}_{d2}\,{\bf q}^{\dagger}_{r2}\right)
+ \left({\bf q}_{r1}\,{\bf q}^{\dagger}_{d1}+{\bf q}_{d1}\,{\bf q}^{\dagger}_{r1}\right)\left( \varrho^2_{r2}\,+\,\varrho^2_{d2}\right)\right\}\,\varepsilon \\
&=\left\{\varrho^2_{r1}\,\varrho^2_{r2}+\varrho^2_{r1}\,\varrho^2_{d2}+\varrho^2_{d1}\,\varrho^2_{r2}+\varrho^2_{d1}\,\varrho^2_{d2} \,+\,\left({\bf q}_{r1}\,{\bf q}^{\dagger}_{d1}+{\bf q}_{d1}\,{\bf q}^{\dagger}_{r1}\right)\left({\bf q}_{r2}\,{\bf q}^{\dagger}_{d2}+{\bf q}_{d2}\,{\bf q}^{\dagger}_{r2}\right)\right\} \notag \\
&\;\;\;\;\;\;\;\;\;\;+\left\{\left( \varrho^2_{r1}\,+\,\varrho^2_{d1}\right) \left({\bf q}_{r2}\,{\bf q}^{\dagger}_{d2}+{\bf q}_{d2}\,{\bf q}^{\dagger}_{r2}\right)
+\left( \varrho^2_{r2}\,+\,\varrho^2_{d2}\right) \left({\bf q}_{r1}\,{\bf q}^{\dagger}_{d1}+{\bf q}_{d1}\,{\bf q}^{\dagger}_{r1}\right)\right\}\,\varepsilon\,, \label{B4}
\end{align}
where I have used ${\bf q}_{r1}\,{\bf q}^{\dagger}_{r1}=\varrho^2_{r1}$, ${\bf q}_{d1}\,{\bf q}^{\dagger}_{d1}=\varrho^2_{d1}$, {\it etc}. Recalling from (\ref{14a}) that the quantities such as ${\bf q}_{r1}\,{\bf q}^{\dagger}_{d1}+{\bf q}_{d1}\,{\bf q}^{\dagger}_{r1}$ are scalar quantities, it is easy to see that the above product (\ref{B4}) also resembles a split complex number similar to ${\mathbb Q}_{z}\,{\mathbb Q}^{\dagger}_{z}$ in (\ref{10c}).

More importantly, the right-hand sides of (\ref{B4}) and (\ref{B8}) are identical. We have thus proved that, although norms in ${{\cal K}^{\lambda}}$ resemble split complex numbers \cite{Dray,Sobczyk} similarly to the product ${\mathbb Q}_{z}\,{\mathbb Q}^{\dagger}_{z}$ in (\ref{10c}) rather than scalar numbers, the composition law (\ref{5}) continues to hold for the algebra ${{\cal K}^{\lambda}}$:
\begin{equation}
||{\mathbb Q}_{z1}\,{\mathbb Q}_{z2}||^2 = ||{\mathbb Q}_{z1}||^2\,||{\mathbb Q}_{z2}||^2. \label{comppp}
\end{equation}
Moreover, the square-root operation can be meaningfully applied to (\ref{5}) by treating the quantity appearing in (\ref{B4}) and (\ref{B8}) as a ``pseudo-complex'' or hyperbolic number $a+\varepsilon\,b$, with $a$ and $b$ being scalars:
\begin{align}
\sqrt{a+\varepsilon\,b\,}
&=\sqrt{\rho\,\{\cosh{(\phi)}+\varepsilon\,\sinh{(\phi)}\}} \\
&=\sqrt{\rho\,\exp{(\varepsilon\,\phi)}} \\ 
&=\sqrt{\rho\,}\,\exp{\left(\varepsilon\,\frac{\phi}{2}\right)} \\
&=\sqrt{\rho\,}\left\{\cosh{\left(\frac{\phi}{2}\right)}+\varepsilon\,\sinh{\left(\frac{\phi}{2}\right)}\right\}.
\end{align}
For details see Sobczyk \cite{Sobczyk}. This allows us to write the composition law (\ref{comppp}) in the square-root form as the norm relation (\ref{no-5}):
\begin{equation}
||{\mathbb Q}_{z1}\,{\mathbb Q}_{z2}|| = ||{\mathbb Q}_{z1}||\,||{\mathbb Q}_{z2}||.
\end{equation}

\subsection{Proof of Lemma~\ref{L3}} \label{Appen2}

To prove this lemma, we begin by recognizing that the algebra ${{\cal K}^{\lambda}}$ remains closed under multiplication. Suppose ${X}$ and ${Y}$ are two multivectors in ${{\cal K}^{\lambda}}$. Then they can be expanded in the graded basis of ${{\cal K}^{\lambda}}$ as:
\begin{equation}
{X}=\,x_0+x_1\,\lambda{\bf e}_x{\bf e}_y+x_2\,\lambda{\bf e}_z{\bf e}_x+x_3\,\lambda{\bf e}_y{\bf e}_z+x_4\,\lambda{\bf e}_x{\bf e}_{\infty}+x_5\,\lambda{\bf e}_y{\bf e}_{\infty}+x_6\,\lambda{\bf e}_z{\bf e}_{\infty}+x_7\,\lambda I_3{\bf e}_{\infty} \label{X}
\end{equation}
and
\begin{equation}
{Y}=\,y_0+y_1\,\lambda{\bf e}_x{\bf e}_y+y_2\,\lambda{\bf e}_z{\bf e}_x+y_3\,\lambda{\bf e}_y{\bf e}_z+y_4\,\lambda{\bf e}_x{\bf e}_{\infty}+y_5\,\lambda{\bf e}_y{\bf e}_{\infty}+y_6\,\lambda{\bf e}_z{\bf e}_{\infty}+y_7\,\lambda I_3{\bf e}_{\infty}\,.\label{Y}
\end{equation}
Using the multiplication table for ${{\cal K}^{\lambda}}$ (Table \ref{T+1}\!\!\!), it is then easy to verify that their product ${{Z}={X}{Y}}$ also belongs to ${{\cal K}^{\lambda}}$:
\begin{align}
{Z}&=\,z_0+z_1\,\lambda{\bf e}_x{\bf e}_y+z_2\,\lambda{\bf e}_z{\bf e}_x+z_3\,\lambda{\bf e}_y{\bf e}_z+z_4\,\lambda{\bf e}_x{\bf e}_{\infty}+z_5\,\lambda{\bf e}_y{\bf e}_{\infty}+z_6\,\lambda{\bf e}_z{\bf e}_{\infty}+z_7\,\lambda I_3{\bf e}_{\infty} \notag \\
&={X}{Y}\in{\cal K}^{\lambda}. \label{Z}
\end{align}
This property can be verified also by writing the two distinct elements of ${{\cal K}^{\lambda}}$ in the form
\begin{equation}
{\mathbb Q}_{z1} =\,{\bf q}_{r1} + {\bf q}_{d1}\,\varepsilon\;\;\;\text{and}\;\;\;{\mathbb Q}_{z2} =\,{\bf q}_{r2} + {\bf q}_{d2}\,\varepsilon\,. \label{nobell}
\end{equation}
Then, using $\varepsilon^2=1$ and the fact that $\varepsilon$ commutes with every element of ${\cal K}^{\lambda}$, their product works out to be
\begin{equation}
{\mathbb Q}_{z3}:=
{\mathbb Q}_{z1}\,{\mathbb Q}_{z2}\,=\,\left({\bf q}_{r1}\,{\bf q}_{r2}\,+\,{\bf q}_{d1}\,{\bf q}_{d2}\right)\,+\,\left({\bf q}_{r1}\,{\bf q}_{d2}\,+\,{\bf q}_{d1}\,{\bf q}_{r2}\right)\,\varepsilon\,. \label{b33}
\end{equation}
But since $\left({\bf q}_{r1}\,{\bf q}_{r2}+{\bf q}_{d1}\,{\bf q}_{d2}\right)$ and $\left({\bf q}_{r1}\,{\bf q}_{d2}+{\bf q}_{d1}\,{\bf q}_{r2}\right)$ are both quaternions, ${\mathbb Q}_{z3}={\mathbb Q}_{z1}\,{\mathbb Q}_{z2}\in{\cal K}^{\lambda}$.

The next question is whether the above property is satisfied by those multivectors of ${{\cal K}^{\lambda}}$ that satisfy the orthogonality condition $\sigma_c^2={\bf q}_{r}\,{\bf q}^{\dagger}_{d}+{\bf q}_{d}\,{\bf q}^{\dagger}_{r}=0$ imposed on them for the purpose of normalization. It turns out that the orthogonality of the quaternions ${{\bf q}_{r}}$ and ${{\bf q}_{d}}$ in ${\mathbb Q}_z$ is preserved under multiplication of the elements of ${\cal K}^{\lambda}$. To prove this, suppose the orthogonality conditions 
\begin{equation}
{\bf q}_{r1}\,{\bf q}^{\dagger}_{d1}+{\bf q}_{d1}\,{\bf q}^{\dagger}_{r1}=0\;\;\;\text{and}\;\;\;{\bf q}_{r2}\,{\bf q}^{\dagger}_{d2}+{\bf q}_{d2}\,{\bf q}^{\dagger}_{r2}=0 \label{B666}
\end{equation}
are satisfied for the two distinct elements considered in (\ref{nobell}). If we now define the following two quaternions,
\begin{equation}
{\bf q}_{r3}\,:=\,\left({\bf q}_{r1}\,{\bf q}_{r2}\,+\,{\bf q}_{d1}\,{\bf q}_{d2}\right)\;\;\;\text{and}\;\;\;
{\bf q}_{d3}\,:=\,\left({\bf q}_{r1}\,{\bf q}_{d2}\,+\,{\bf q}_{d1}\,{\bf q}_{r2}\right),
\end{equation}
then the product (\ref{b33}) of the multivectors considered in (\ref{nobell}) can be expressed as the following third multivector within ${\cal K}^{\lambda}$: 
\begin{equation}
{\mathbb Q}_{z3}=\,{\bf q}_{r3} + {\bf q}_{d3}\,\varepsilon\,.
\label{notso}
\end{equation}
It is now easy to prove that the quaternions ${{\bf q}_{r3}}$ and ${{\bf q}_{d3}}$ are also orthogonal, or equivalently, that ${{\bf q}_{r3}\,{\bf q}^{\dagger}_{d3}+{\bf q}_{d3}\,{\bf q}^{\dagger}_{r3}=0}$:
\begin{align}
{\bf q}_{r3}\,{\bf q}^{\dagger}_{d3}+{\bf q}_{d3}\,{\bf q}^{\dagger}_{r3}\,&=\,
\left({\bf q}_{r1}\,{\bf q}_{r2}\,+\,{\bf q}_{d1}\,{\bf q}_{d2}\right)
\left({\bf q}_{r1}\,{\bf q}_{d2}\,+\,{\bf q}_{d1}\,{\bf q}_{r2}\right)^{\dagger}+\,
\left({\bf q}_{r1}\,{\bf q}_{d2}\,+\,{\bf q}_{d1}\,{\bf q}_{r2}\right)
\left({\bf q}_{r1}\,{\bf q}_{r2}\,+\,{\bf q}_{d1}\,{\bf q}_{d2}\right)^{\dagger} \\
&=\,
\left({\bf q}_{r1}\,{\bf q}_{r2}\,+\,{\bf q}_{d1}\,{\bf q}_{d2}\right)
\left({\bf q}_{d2}^{\dagger}\,{\bf q}_{r1}^{\dagger}\,+\,{\bf q}_{r2}^{\dagger}\,{\bf q}_{d1}^{\dagger}\right)+\,
\left({\bf q}_{r1}\,{\bf q}_{d2}\,+\,{\bf q}_{d1}\,{\bf q}_{r2}\right)
\left({\bf q}_{r2}^{\dagger}\,{\bf q}_{r1}^{\dagger}\,+\,{\bf q}_{d2}^{\dagger}\,{\bf q}_{d1}^{\dagger}\right) \\
&=\,
{\bf q}_{r1}\,{\bf q}_{r2}\,{\bf q}_{d2}^{\dagger}\,{\bf q}_{r1}^{\dagger}\,+\,{\bf q}_{r1}\,{\bf q}_{r2}\,{\bf q}_{r2}^{\dagger}\,{\bf q}_{d1}^{\dagger}\,+\,{\bf q}_{d1}\,{\bf q}_{d2}\,{\bf q}_{d2}^{\dagger}\,{\bf q}_{r1}^{\dagger}\,+\,{\bf q}_{d1}\,{\bf q}_{d2}\,{\bf q}_{r2}^{\dagger}\,{\bf q}_{d1}^{\dagger} \notag \\
&\;\;\;\;\;\;\;\;+
\,{\bf q}_{r1}\,{\bf q}_{d2}\,{\bf q}_{r2}^{\dagger}\,{\bf q}_{r1}^{\dagger}\,+\,{\bf q}_{r1}\,{\bf q}_{d2}\,{\bf q}_{d2}^{\dagger}\,{\bf q}_{d1}^{\dagger}\,+\,{\bf q}_{d1}\,{\bf q}_{r2}\,{\bf q}_{r2}^{\dagger}\,{\bf q}_{r1}^{\dagger}\,+\,{\bf q}_{d1}\,{\bf q}_{r2}\,{\bf q}_{d2}^{\dagger}\,{\bf q}_{d1}^{\dagger} \\
&=\,{\bf q}_{r1}\left\{{\bf q}_{r2}\,{\bf q}_{d2}^{\dagger}+{\bf q}_{d2}\,{\bf q}_{r2}^{\dagger}\right\}{\bf q}_{r1}^{\dagger}\,+\,\varrho_{r2}^2\left\{{\bf q}_{r1}\,{\bf q}_{d1}^{\dagger}+{\bf q}_{d1}\,{\bf q}_{r1}^{\dagger}\right\} \notag \\
&\;\;\;\;\;\;\;\;+\,{\bf q}_{d1}\left\{{\bf q}_{r2}\,{\bf q}_{d2}^{\dagger}+{\bf q}_{d2}\,{\bf q}_{r2}^{\dagger}\right\}{\bf q}_{d1}^{\dagger}\,+\,\varrho_{d2}^2\left\{{\bf q}_{r1}\,{\bf q}_{d1}^{\dagger}+{\bf q}_{d1}\,{\bf q}_{r1}^{\dagger}\right\} \label{c9}\\ 
&=\,\left(\varrho_{r1}^2+\varrho_{d1}^2\right)\left\{{\bf q}_{r2}\,{\bf q}_{d2}^{\dagger}+{\bf q}_{d2}\,{\bf q}_{r2}^{\dagger}\right\}\,+\,\left(\varrho_{r2}^2+\varrho_{d2}^2\right)\left\{{\bf q}_{r1}\,{\bf q}_{d1}^{\dagger}+{\bf q}_{d1}\,{\bf q}_{r1}^{\dagger}\right\} \label{c10}\\
&=\,0.
\end{align}
Here Eq.~(\ref{c10}) follows from Eq.~(\ref{c9}) because, as shown in Eq.~(\ref{normcon}), the quantities such as ${\bf q}_{r1}\,{\bf q}_{d1}^{\dagger}+{\bf q}_{d1}\,{\bf q}_{r1}^{\dagger}$ are purely scalar quantities, and therefore we can use 
${\bf q}_{r1}\,{\bf q}^{\dagger}_{r1}=\varrho^2_{r1}$, ${\bf q}_{d1}\,{\bf q}^{\dagger}_{d1}=\varrho^2_{d1}$, {\it etc}., as before, to reduce Eq.~(\ref{c9}) to
Eq.~(\ref{c10}). Then, the orthogonality conditions ${\bf q}_{r1}\,{\bf q}^{\dagger}_{d1}+{\bf q}_{d1}\,{\bf q}^{\dagger}_{r1}=0$ and ${\bf q}_{r2}\,{\bf q}^{\dagger}_{d2}+{\bf q}_{d2}\,{\bf q}^{\dagger}_{r2}=0$ from (\ref{B666}) reduces the RHS of Eq.~(\ref{c10}) to zero.\break Consequently, the 7-sphere defined in (\ref{sevsp}) remains closed under multiplication, analogously to the octonionic 7-sphere.

\subsection{Proof of Lemma~\ref{L4}} \label{Appen3}

Let ${U}$ and ${V}$ be any two multivectors in ${\cal K}^{\lambda}$ defined by (\ref{RepR}), subject to the normalization condition $\sigma_c^2={\bf q}_{r}{\bf q}^{\dagger}_{d}+{\bf q}_{d}{\bf q}^{\dagger}_{r}=0$. Then their norms $||{U}||$ and $||{V}||$ are scalar numbers. Therefore, the product of their norms $||{U}||\,||{V}||$ can vanish if and only if either $||{U}||=0$ or $||{V}||=0$. And the positive definiteness (\ref{pd}) of the norms implies that the latter is possible if and only if either ${U}=0$ or ${V}=0$. In other words, if both ${U}\not=0$ and ${V}\not=0$, then $||{U}||\,||{V}||\not=0$, unlike that in (\ref{51-n}). The latter result, namely ${||{X}||\,||{Y}||=0}$ for ${X}\not=0$ and ${Y}\not=0$, occurred because of the hyperbolic norm values $||{X}||=\sqrt{2\,(1-\varepsilon)}\,$ and $||{Y}||=\sqrt{2\,(1+\varepsilon)}$, which is not possible for the scalar norm values of $||{U}||$ and $||{V}||$. But that implies ${||{U}{V}||\not=0}$ because, according to (\ref{normfinal}), $||{U}||\,||{V}||=||{U}{V}||$. The positive definiteness (\ref{pd}) of the norms then implies ${{U}{V}\not=0}$. Thus for arbitrary  ${U}\not=0$ and ${V}\not=0$ we have proved that ${{U}{V}\not=0}$. Therefore the normed algebra ${\cal K}^{\lambda}$ in (\ref{RepR}) subject to (\ref{normcon}) is a division algebra.

\subsection{Proof of Eq. (\ref{limitproduct})} \label{Appen5}

To prove the equality displayed in Eq.~(\ref{limitproduct}), we begin with the equalities (\ref{84-nom}) = (\ref{79-nmn}) and (\ref{90-nom}) = (\ref{85-nmn}), which are as follows:
\begin{equation}
\lim_{{\mathbf s}_1\,\rightarrow\,\mu_1{\mathbf a}}\left\{\,+\,{\mathbf q}(\eta_{{\mathbf a}{\mathbf s}_1},\,{\mathbf r}_1)\right\} = \lim_{{\mathbf s}_1\,\rightarrow\,\mu_1{\mathbf a}}\left\{-\,{\mathbf D}({\mathbf a})\,{\mathbf L}({\mathbf s}_1)\right\} \label{sim-1}
\end{equation}
and
\begin{equation}
\lim_{{\mathbf s}_2\,\rightarrow\,\mu_2{\mathbf b}}\left\{\,-\,{\mathbf q}(\eta_{{\mathbf s}_2{\mathbf b}},\,{\mathbf r}_2)\right\}=\lim_{{\mathbf s}_2\,\rightarrow\,\mu_2{\mathbf b}}\left\{+\,{\mathbf L}({\mathbf s}_2)\,{\mathbf D}({\mathbf b})\right\}. \label{sim-2}
\end{equation}
Using these equalities, the product of limits on the left-hand side of (\ref{limitproduct}) can be simplified as follows:
\begin{align}
\left[\lim_{{\mathbf s}_1\,\rightarrow\,\mu_1{\mathbf a}}\left\{+\,{\mathbf q}(\eta_{{\mathbf a}{\mathbf s}_1},\,{\mathbf r}_1)\right\}\right]\left[\lim_{{\mathbf s}_2\,\rightarrow\,\mu_2{\mathbf b}}\left\{-\,{\mathbf q}(\eta_{{\mathbf s}_2{\mathbf b}},\,{\mathbf r}_2)\right\}\right]
&=\,\bigg[\lim_{{\mathbf s}_1\,\rightarrow\,\mu_1{\mathbf a}}\left\{-\,{\mathbf D}({\mathbf a})\,{\mathbf L}({\mathbf s}_1)\right\}\bigg]\left[\lim_{{\mathbf s}_2\,\rightarrow\,\mu_2{\mathbf b}}\left\{+\,{\mathbf L}({\mathbf s}_2)\,{\mathbf D}({\mathbf b})\right\}\right] \\
&=\,\bigg[\lim_{{\mathbf s}_1\,\rightarrow\,\mu_1{\mathbf a}}\left\{-I_3{\mathbf a}\,I_3{\mathbf s}_1\right\}\bigg]\left[\lim_{{\mathbf s}_2\,\rightarrow\,\mu_2{\mathbf b}}\left\{+I_3{\mathbf s}_2\,I_3{\mathbf b}\right\}\right] \\
&=\,\bigg[\lim_{{\mathbf s}_1\,\rightarrow\,\mu_1{\mathbf a}}\left\{{\mathbf a}\,{\mathbf s}_1\right\}\bigg]\left[\lim_{{\mathbf s}_2\,\rightarrow\,\mu_2{\mathbf b}}\left\{-\,{\mathbf s}_2\,{\mathbf b}\right\}\right] \\
&=\,\left[\mu_1{\mathbf a}\,{\mathbf a}\right]\left[-\mu_2{\mathbf b}\,{\mathbf b}\right] \\
&=-\mu_1\mu_2\,, \label{id1}
\end{align}
because all vectors involved in the above equations are unit vectors and I have used $(I_3)^2=-1$ and the fact that the pseudoscalar $I_3$ commutes with all other elements of $\mathrm{Cl}_{3,0}$. Similarly, using (\ref{sim-1}) and (\ref{sim-2}) the right-hand side of (\ref{limitproduct}) can be simplified to
\begin{align}
\lim_{\substack{{\mathbf s}_1\,\rightarrow\,\mu_1{\mathbf a} \\ {\mathbf s}_2\,\rightarrow\,\mu_2{\mathbf b}}}\Big\{-{\mathbf q}(\eta_{{\mathbf a}{\mathbf s}_1},\,{\mathbf r}_1)\,{\mathbf q}(\eta_{{\mathbf s}_2{\mathbf b}},\,{\mathbf r}_2)\Big\}
&=\lim_{\substack{{\mathbf s}_1\,\rightarrow\,\mu_1{\mathbf a} \\ {\mathbf s}_2\,\rightarrow\,\mu_2{\mathbf b}}}\left[   \left\{-\,{\mathbf D}({\mathbf a})\,{\mathbf L}({\mathbf s}_1)\right\}\left\{+\,{\mathbf L}({\mathbf s}_2)\,{\mathbf D}({\mathbf b})\right\}\right] \\
&=\lim_{\substack{{\mathbf s}_1\,\rightarrow\,\mu_1{\mathbf a} \\ {\mathbf s}_2\,\rightarrow\,\mu_2{\mathbf b}}}\left[   \left\{-I_3{\mathbf a}\,I_3{\mathbf s}_1\right\}\left\{+I_3{\mathbf s}_2\,I_3{\mathbf b})\right\}\right] \\
&=\lim_{\substack{{\mathbf s}_1\,\rightarrow\,\mu_1{\mathbf a} \\ {\mathbf s}_2\,\rightarrow\,\mu_2{\mathbf b}}}\left[   \left\{{\mathbf a}\,{\mathbf s}_1\right\}\left\{-\,{\mathbf s}_2{\mathbf b})\right\}\right] \\
&=\lim_{\substack{{\mathbf s}_1\,\rightarrow\,\mu_1{\mathbf a} \\ {\mathbf s}_2\,\rightarrow\,\mu_2{\mathbf b}}}\left[-\,{\mathbf a}\,{\mathbf s}_1\,{\mathbf s}_2{\mathbf b}\right] \\
&=\left[-\mu_1{\mathbf a}\,{\mathbf a}\,\mu_2{\mathbf b}\,{\mathbf b}\right] \\
&=-\mu_1\mu_2. \label{id2}
\end{align}
Since the right-hand sides of (\ref{id1}) and (\ref{id2}) are identical, ``the product of limits = limits of product'' holds, as claimed in (\ref{limitproduct}).  

\subsection{Proof of Eq. (\ref{6forappen})} \label{Appen4} 

Let us begin by expanding the two quaternions appearing on the left-hand side of Eq.~(\ref{q-eq-appen}) in the standard form as follows:
\begin{equation}
{\mathbf q}(\eta_{{\mathbf a}{\mathbf s}_1},\,{\mathbf r}_{1})=\cos(\eta_{{\mathbf a}{\mathbf s}_1}) + (I_3{\mathbf r}_1)\,\sin(\eta_{{\mathbf a}{\mathbf s}_1}) \label{9-nn}
\end{equation}
and
\begin{equation}
{\mathbf q}(\eta_{{\mathbf s}_2{\mathbf b}},\,{\mathbf r}_{2})=\cos(\eta_{{\mathbf s}_2{\mathbf b}}) + (I_3{\mathbf r}_2)\,\sin(\eta_{{\mathbf s}_2{\mathbf b}}). \label{8-nn}
\end{equation}
The quaternion ${\mathbf q}(\eta_{{\mathbf u}{\mathbf v}},\,{\mathbf r}_{0})$ on the right-hand side of (\ref{q-eq-appen}) is then the product of these quaternions and can be evaluated as follows:
\begin{align}
{\mathbf q}(\eta_{{\mathbf u}{\mathbf v}},\,{\mathbf r}_{0})&=\{{\mathbf q}(\eta_{{\mathbf a}{\mathbf s}_1},\,{\mathbf r}_{1})\}\{{\mathbf q}(\eta_{{\mathbf s}_2{\mathbf b}},\,{\mathbf r}_{2})\} \label{115-n} \\
&=\{\cos(\eta_{{\mathbf a}{\mathbf s}_1}) + (I_3{\mathbf r}_1)\,\sin(\eta_{{\mathbf a}{\mathbf s}_1})\} \{\cos(\eta_{{\mathbf s}_2{\mathbf b}}) + (I_3{\mathbf r}_2)\,\sin(\eta_{{\mathbf s}_2{\mathbf b}})\} \\
&=\cos(\eta_{{\mathbf a}{\mathbf s}_1})\cos(\eta_{{\mathbf s}_2{\mathbf b}}) + (I_3{\mathbf r}_2)\cos(\eta_{{\mathbf a}{\mathbf s}_1})\sin(\eta_{{\mathbf s}_2{\mathbf b}}) + (I_3{\mathbf r}_1)\cos(\eta_{{\mathbf s}_2{\mathbf b}})\sin(\eta_{{\mathbf a}{\mathbf s}_1}) + (I_3{\mathbf r}_1)(I_3{\mathbf r}_2)\sin(\eta_{{\mathbf a}{\mathbf s}_1})\sin(\eta_{{\mathbf s}_2{\mathbf b}}).\label{99-n}
\end{align}
Here $(I_3{\mathbf r}_1)(I_3{\mathbf r}_2)$ appearing in the last term is a geometric product of two bivectors, which can be expanded as follows:
\begin{align}
(I_3{\mathbf r}_1)(I_3{\mathbf r}_2)&= -{\mathbf r}_1\,{\mathbf r}_2 \\
&= -{\mathbf r}_1\cdot{\mathbf r}_2-{\mathbf r}_1\wedge{\mathbf r}_2 \\
&= -{\mathbf r}_1\cdot{\mathbf r}_2-I_3({\mathbf r}_1\times{\mathbf r}_2), \label{118-n}
\end{align}
where $I_3^2=({\bf e}_x{\bf e}_y{\bf e}_z)({\bf e}_x{\bf e}_y{\bf e}_z)=-1$ is used. Substituting the right-hand side of (\ref{118-n}) into (\ref{99-n}), the latter simplifies to
\begin{align}
{\mathbf q}(\eta_{{\mathbf u}{\mathbf v}},\,{\mathbf r}_{0})=\cos(\eta_{{\mathbf a}{\mathbf s}_1})\cos(\eta_{{\mathbf s}_2{\mathbf b}}) + (I_3{\mathbf r}_2)\cos(\eta_{{\mathbf a}{\mathbf s}_1})&\sin(\eta_{{\mathbf s}_2{\mathbf b}}) +(I_3{\mathbf r}_1)\cos(\eta_{{\mathbf s}_2{\mathbf b}})\sin(\eta_{{\mathbf a}{\mathbf s}_1}) \notag \\
&-({\mathbf r}_1\cdot{\mathbf r}_2)\sin(\eta_{{\mathbf a}{\mathbf s}_1})\sin(\eta_{{\mathbf s}_2{\mathbf b}}) - \{I_3({\mathbf r}_1\times{\mathbf r}_2)\}\sin(\eta_{{\mathbf a}{\mathbf s}_1})\sin(\eta_{{\mathbf s}_2{\mathbf b}}). \label{102-n}
\end{align}
Next, we wish to express the quaternion ${\mathbf q}(\eta_{{\mathbf u}{\mathbf v}},\,{\mathbf r}_{0})$ in the standard form, with $\eta_{{\mathbf u}{\mathbf v}}$ being half the rotation angle of the quaternion:
\begin{equation}
{\mathbf q}(\eta_{{\mathbf u}{\mathbf v}},\,{\mathbf r}_{0})=\cos(\eta_{{\mathbf u}{\mathbf v}}) + (I_3{\mathbf r}_0)\,\sin(\eta_{{\mathbf u}{\mathbf v}}). \label{101-n}
\end{equation}
This can be done by collecting the sum of all scalar parts from (\ref{102-n}) and identifying it with $\cos(\eta_{{\mathbf u}{\mathbf v}})$, and collecting the sum of all bivector parts from (\ref{102-n}) and identifying it with $I_3\{{\mathbf r}_0\,\sin(\eta_{{\mathbf u}{\mathbf v}})\}$:
\begin{equation}
\cos(\eta_{{\mathbf u}{\mathbf v}})=\{\cos(\eta_{{\mathbf a}{\mathbf s}_1})\cos(\eta_{{\mathbf s}_2{\mathbf b}}) -({\mathbf r}_1\cdot{\mathbf r}_2)\sin(\eta_{{\mathbf a}{\mathbf s}_1})\sin(\eta_{{\mathbf s}_2{\mathbf b}})\} \label{102-nnm}
\end{equation}
and
\begin{equation}
I_3\{{\mathbf r}_0\,\sin(\eta_{{\mathbf u}{\mathbf v}})\}=I_3\{{\mathbf r}_1\cos(\eta_{{\mathbf s}_2{\mathbf b}})\sin(\eta_{{\mathbf a}{\mathbf s}_1}) + {\mathbf r}_2\cos(\eta_{{\mathbf a}{\mathbf s}_1})\sin(\eta_{{\mathbf s}_2{\mathbf b}}) - ({\mathbf r}_1\times{\mathbf r}_2)\,\sin(\eta_{{\mathbf a}{\mathbf s}_1})\sin(\eta_{{\mathbf s}_2{\mathbf b}})\}. \label{103-n}
\end{equation}
Using the definitions in (\ref{rot1}) and (\ref{rot2}) of the axis vectors ${\mathbf r}_1$ and ${\mathbf r}_2$, together with $||({\mathbf a}\times{\mathbf s}_1)||=\sin(\eta_{{\mathbf a}{\mathbf s}_1})$ and $||({\mathbf s}_2\times{\mathbf b})||=\sin(\eta_{{\mathbf s}_2{\mathbf b}})$ from the normalizations of cross products, (\ref{102-nnm}) simplifies to
\begin{equation}
\cos(\eta_{{\mathbf u}{\mathbf v}})=\{({\mathbf a}\cdot{\mathbf s}_1)({\mathbf s}_2\cdot{\mathbf b})-({\mathbf a}\times{\mathbf s}_1)\cdot({\mathbf s}_2\times{\mathbf b})\}.
\end{equation}
This can be further simplified by using the vector identity
\begin{equation}
({\mathbf a}\times{\mathbf s}_1)\cdot({\mathbf s}_2\times{\mathbf b})=({\mathbf a}\cdot{\mathbf s}_2)({\mathbf s}_1\cdot{\mathbf b})-
({\mathbf a}\cdot{\mathbf b})({\mathbf s}_1\cdot{\mathbf s}_2).
\end{equation}
Consequently, we arrive at the angle $\eta_{{\mathbf u}{\mathbf v}}$ expressed in Eq.~(\ref{38-22}) of Section~\ref{Sec-5}:
\begin{equation}
\eta_{{\mathbf u}{\mathbf v}}({\mathbf s}_1,\,{\mathbf s}_2)=\cos^{-1}\big\{({\mathbf a}\cdot{\mathbf s}_1)({\mathbf s}_2\cdot{\mathbf b}) - ({\mathbf a}\cdot{\mathbf s}_2)({\mathbf s}_1\cdot{\mathbf b}) + ({\mathbf a}\cdot{\mathbf b})({\mathbf s}_1\cdot{\mathbf s}_2)\big\}. \label{55-nmn}
\end{equation}
On the other hand, solving equation (\ref{103-n}) for ${\mathbf r}_0({\mathbf s}_1,\,{\mathbf s}_2)$ gives   
\begin{equation}
{\mathbf r}_{0}({\mathbf s}_1,\,{\mathbf s}_2)=\frac{{\mathbf r}_1 \cos\left(\,\eta_{{\mathbf s}_2{\mathbf b}}\right)\sin\left(\,\eta_{{\mathbf a}{\mathbf s}_1}\right) + {\mathbf r}_2\cos\left(\,\eta_{{\mathbf a}{\mathbf s}_1}\right)\sin\left(\,\eta_{{\mathbf s}_2{\mathbf b}}\right) - ({\mathbf r}_1\times{\mathbf r}_2) \sin\left(\,\eta_{{\mathbf a}{\mathbf s}_1}\right)\sin\left(\,\eta_{{\mathbf s}_2{\mathbf b}}\right)}{\sin\left(\,\eta_{{\mathbf a}{\mathbf b}}\right)}. \label{107-n}
\end{equation}
Again using the definitions (\ref{rot1}) and (\ref{rot2}) of the axis vectors ${\mathbf r}_1$ and ${\mathbf r}_2$ and $||({\mathbf a}\times{\mathbf s}_1)||=\sin(\eta_{{\mathbf a}{\mathbf s}_1})$ and $||({\mathbf s}_2\times{\mathbf b})||=\sin(\eta_{{\mathbf s}_2{\mathbf b}})$ from the normalizations of the cross products, (\ref{107-n}) simplifies to
\begin{equation}
{\mathbf r}_{0}({\mathbf s}_1,\,{\mathbf s}_2)=\frac{({\mathbf a}\cdot{\mathbf s}_1)({\mathbf s}_2\times{\mathbf b})+({\mathbf s}_2\cdot{\mathbf b})({\mathbf a}\times{\mathbf s}_1)-({\mathbf a}\times{\mathbf s}_1)\times({\mathbf s}_2\times{\mathbf b})}{\sin\left(\,\eta_{{\mathbf a}{\mathbf b}}\right)},
\end{equation}
which is Eq.~(\ref{101exam}) of Section~\ref{Sec-5}. With the above explicit expressions of ${\mathbf r}_{0}({\mathbf s}_1,\,{\mathbf s}_2)$, for ${\mathbf s}_1=\,{\mathbf s}_2$ the quaternion (\ref{101-n}) reduces to
\begin{equation}
{\mathbf q}(\eta_{{\mathbf a}{\mathbf b}},\,{\mathbf r}_{0})=\cos(\eta_{{\mathbf a}{\mathbf b}}) + (I_3{\mathbf r}_0)\,\sin(\eta_{{\mathbf a}{\mathbf b}}),
\end{equation}
because, as evident from (\ref{55-nmn}), the angle $\eta_{{\mathbf u}{\mathbf v}}$ reduces to $\eta_{{\mathbf a}{\mathbf b}}$ if ${\mathbf s}_1=\,{\mathbf s}_2$ holds, as required by conservation of angular momentum. We thus see that the product ${\mathbf q}(\eta_{{\mathbf a}{\mathbf s}_1},\,{\mathbf r}_{1})\,{\mathbf q}(\eta_{{\mathbf s}_2{\mathbf b}},\,{\mathbf r}_{2})$ in Eq.~(\ref{5forappen}) reduces to the quaternion ${\mathbf q}(\eta_{{\mathbf a}{\mathbf b}},\,{\mathbf r}_{0})$ in Eq.~(\ref{6forappen}) for ${\mathbf s}_1=\,{\mathbf s}_2$.

\end{document}